\documentclass{fairmeta}

\usepackage[utf8]{inputenc}
\usepackage{url}
\usepackage{amsfonts}
\usepackage{nicefrac}
\usepackage{colortbl}

\usepackage{tikz}
\usepackage{adjustbox}
\usepackage{array}
\usetikzlibrary{matrix,fit}
\usepackage{comment}
\usepackage{float}

\usepackage{algorithm, algpseudocode}

\usepackage{amsmath,amssymb}
\usepackage{amsthm}
\usepackage{bbm}

\usepackage{pifont}
\usepackage{xspace}
\usepackage{enumitem}

\newcommand{\ie}{{i.e}.\@ }

\newcommand{\OURS}{\textsc{CwA}\xspace}
\newcommand{\PROD}{\textsc{CwA-Prod}\xspace}
\newcommand{\FLAT}{\textsc{CwA}\xspace}
\newcommand{\HNSW}{\textsc{CwA-HNSW}\xspace}

\colorlet{ourcol}{blue!8}

\newcommand{\nprobe}{m}
\newcommand{\ntrain}{{\ensuremath{n_{\text{train}}}\xspace}}
\newcommand{\nindex}{{\ensuremath{n_{\text{index}}}\xspace}}

\title{Cluster with Auctions for Vector Search}

\author[1]{Swann Bessa}
\author[1]{Pierre Fernandez}
\author[1]{Gergely Szilvasy}
\author[1]{Matthijs Douze}
\author[1]{Herv\'e J\'egou}

\affiliation[1]{Meta FAIR}

\abstract{%
Large-scale approximate nearest neighbor search commonly relies on partitions for indexing: database vectors are partitioned into clusters, and for each query
a probing function selects the clusters to be scanned.
The query probing function and the database partition are rarely treated as separate entities: most techniques assign queries with the same assignment function as the database vectors, which is suboptimal especially when database and query distributions differ. This paper introduces \textbf{CwA (Cluster with Auctions)}, which addresses this limitation by jointly learning a balanced database partition and a neural probing function. \OURS optimizes search performance directly for the query distribution. It minimizes its objective by alternating two steps: (i) gradient descent on the neural network of the probing function, and (ii) a large-scale combinatorial optimization of the cluster assignment for the database vectors.
We solve the latter with a parallelizable auction algorithm that balances the partition by design. To further scale \OURS, we extend the method to a Cartesian product of clusters that increases the partition's granularity. When database and query distributions differ, \OURS achieves up to 4.7$\times$ throughput over the state-of-the-art at equal recall.
In the in-distribution (ID) setting, even a simple linear probing function trained with \OURS outperforms competing deep neural methods.%
}

\metadata[Correspondence at]{swann@meta.com}

\begin{document}

\maketitle

\begin{figure}[h!]
    \begin{minipage}[t]{0.49\textwidth}
        \includegraphics[width=\linewidth]{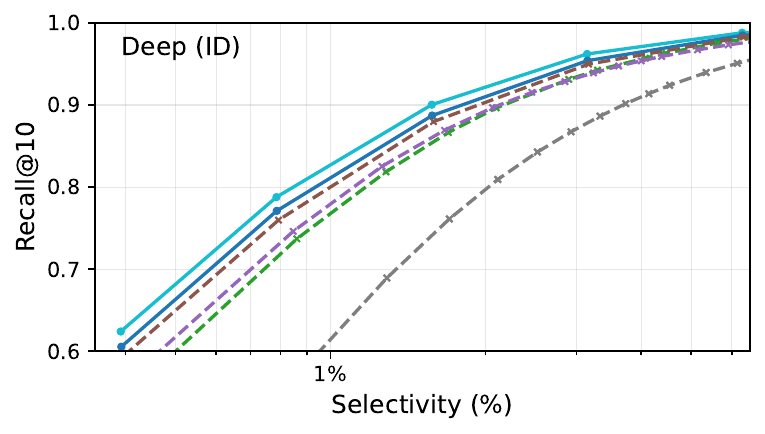}
    \end{minipage}
    \hfill
    \begin{minipage}[t]{0.49\textwidth}
        \includegraphics[width=\linewidth]{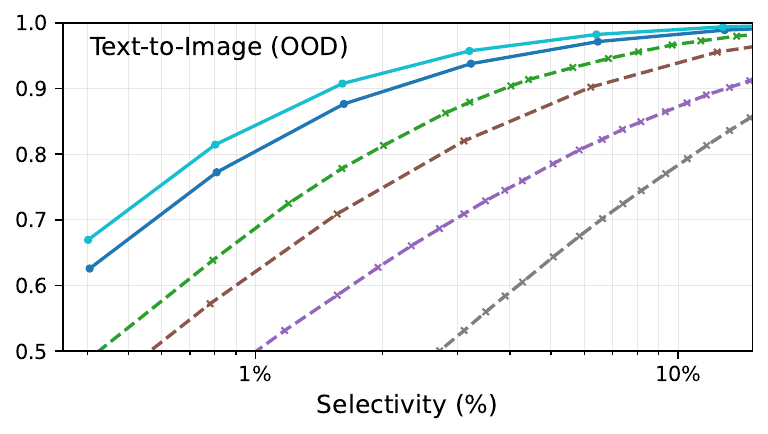}
    \end{minipage}

    \vspace{-4pt}
    \includegraphics[width=1.\textwidth]{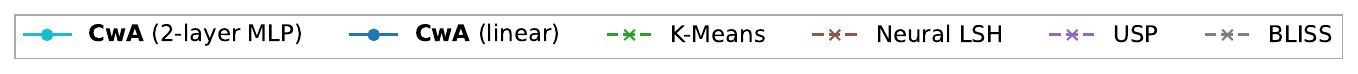}
    \caption{%
        \OURS (ours) vs.\ baselines on $1\text{M}$ vectors, 256 clusters.
        Selectivity is the fraction of database vectors that we compare the queries with (smaller = faster).
        \emph{Left:} Deep dataset (in-distribution). \OURS matches or outperforms all baselines, including deep networks with more parameters.
        \emph{Right:} Text-to-Image dataset (out-of-distribution). The performance gap widens dramatically: \OURS largely outperforms the best baseline K-Means, while other neural approaches from the state of the art underperform.
    \label{fig:flagship_text2image10M}}
\end{figure}

\section{Introduction}
\label{sec:intro}
Nearest neighbor search on high-dimensional vectors is used in a wide range of applications, from retrieval-augmented generation for large language models~\citep{lewis2020retrieval} to reverse image search on learned embeddings~\citep{babenko2016efficient,pizzi2022self} and large-scale deduplication~\citep{douze2025machine}.
At the scales encountered in practice, exact search is prohibitively expensive, motivating the use of approximate nearest neighbor search (ANNS) methods that trade a controlled amount of accuracy for significantly higher throughput.

Partition-based indexing is one of the most widely adopted ANNS methods: a database is split into clusters, and for each query only a subset of these clusters---those most likely to contain its nearest neighbors---is visited. This naturally involves two asymmetric roles: a \textit{partition} performs a discrete assignment of database vectors to clusters, while a \textit{probing function} scores which clusters to search for each query.
The canonical instance, the inverted file index IVF~\citep{sivic2003videogoogle}, relies on K-Means centroids: nearest-centroid assignment is used to split the dataset into clusters and distances to centroids serve as the probing function.

We argue that three properties are crucial for optimal search performance.
First, the partition and the probing function should \textit{not be conflated}. 
In recent neural approaches for IVF~\citep{dong2020learningspacepartitionsnearest,fahim2022unsupervisedspacepartitioningnearest,Gupta2022BLISSAB}, the same model partitions the database vectors and serves as a probing function for the queries. Sharing a single model for these two different tasks unnecessarily constrains the design space.
Second, both the partition and the probing function should be \textit{adapted to the query distribution}.
In the out-of-distribution (OOD) setting, queries and database vectors follow different distributions. 
Therefore, clusters trained only on the database poorly cover the regions queries fall into~\citep{Chen_2024_roargraph}.
Finally, the partition should be \textit{balanced}: size imbalance introduces query latency variance, inflating worst-case latency.

\OURS (\textit{Cluster with Auctions}) addresses all three properties.
It separates the two roles. 
A neural network is trained on the query distribution to predict which clusters contain a query's nearest neighbors.
On the other hand, the partition of the database vectors is obtained by solving a capacitated linear assignment problem with the auction algorithm~\citep{bertsekas1988auction}.
Both steps are repeatedly alternated to decrease a common loss.
The cluster balance is enforced during the auction stage.     

When indexing larger datasets, the most efficient partitions use a finer cluster granularity. 
For datasets of more than 10M vectors, we propose \PROD, a variant that forms a large codebook via a Cartesian product of two smaller ones. 
\PROD scales to hundreds of thousands of clusters with low parameter count and fast inference.

In summary, our contributions are:
\begin{itemize}[leftmargin=2em]
\setlength\itemsep{0em}
    \item A joint optimization of a neural probing function and a balanced partition via alternating optimization, with consistent decrease of a unified query-dependent loss (Section \ref{sec:method}); 
    \item A formulation of the partition update as a capacitated linear assignment problem, solved efficiently on GPU with the auction algorithm;
    \item A product-key variant that scales to very large numbers of clusters;
    \item State-of-the-art results on four large-scale benchmarks in both in-distribution and out-of-distribution settings (see Figure~\ref{fig:flagship_text2image10M} and Section~\ref{sec:experiments}), demonstrating that previous approaches left significant room for improvement in learned partitioning.
\end{itemize}

\section{Background: nearest neighbor search and capacitated linear assignment}
\label{sec:background}

This section introduces the technical prerequisites for the method presented in our paper. 
The main notations are summarized in Appendix~\ref{sec:notation}.

\subsection{Nearest neighbor search}
\label{sec:background-nn}

Given a dataset of d-dimensional vectors $X = (x_{\ell})_{\ell=1,\dots,\nindex} \in \mathbb{R}^d$ and a query $q \in \mathbb{R}^d$, we aim to find the query's k-nearest neighbors (k-NNs) in $X$ according to a given distance metric. 

Denoting $\mathcal{N}_k(q)$ the ground-truth NNs and $\hat{\mathcal{N}}_k(q)$ the NNs returned by an approximate search algorithm, the quality of the result is measured by the \emph{Recall@k} metric: \mbox{$\text{Recall@}k(q) = |\mathcal{N}_k(q) \cap \hat{\mathcal{N}}_k(q)| / k$}, which we aim to maximize throughout this work.

\subsection{Partition-based indexing}
\label{subsec:partition-indexing}

Our approach relies on a \emph{partition} of dataset $X$ into a set of clusters $C$.
We consider $X$ to be fixed; the out-of-sample extension is outside the scope of this work.
Therefore, this partition is represented by a \emph{partitioning function} $h$ mapping each database vector id to its cluster id:
$h: \{1,\dots,\nindex \} \to C$. 
At inference time, given the query $q\in\mathbb{R}^d$, a \emph{probing function} $f_{\theta}(q) \in [0,1]^{|C|} $ yields a probability distribution over clusters and the top-$\nprobe$ outputs of that function are the clusters most likely to contain the k-NNs of $q$. 
This is known as \emph{multi-probe querying}.
Note that $h$ is a mapping based on ids, while $f_\theta$ is computed from vector coordinates. 
The probed clusters are a subset of $X$, over which distances to $q$  are computed to output $\hat{\mathcal{N}}_k(q)$.

\subsection{Capacitated Linear Assignment Problem}
\label{subsec:lin_ass_prob}

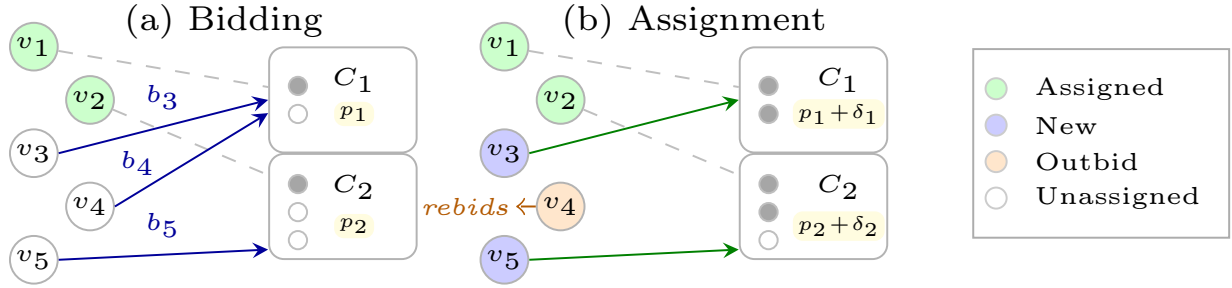
\begin{figure}
    \centering
    \resizebox{1\linewidth}{!}{%
        \begin{tikzpicture}[
    vector/.style={circle, draw=gray!60, thin, minimum size=0.34cm, font=\tiny, inner sep=0pt},
    assigned/.style={vector, fill=green!20},
    unassigned/.style={vector, fill=white},
    loser/.style={vector, fill=orange!20},
    newassigned/.style={vector, fill=blue!20},
    cluster/.style={rectangle, draw=gray!60, thin, rounded corners=3pt,
                    minimum width=1.05cm, minimum height=0.75cm},
    slot/.style={circle, draw=gray!50, thin, minimum size=0.13cm, inner sep=0pt},
    slotfilled/.style={slot, fill=gray!70},
    slotempty/.style={slot, fill=white},
    arrow/.style={->, >=stealth, thin},
    bidarrow/.style={arrow, blue!60!black},
    assignline/.style={-, gray!50, thin, dashed},
    pricelabel/.style={font=\tiny, fill=yellow!15, rounded corners=2pt, inner sep=1pt}
]
\pgfdeclarelayer{background}
\pgfsetlayers{background,main}

\begin{scope}[local bounding box=panel1]

    \node[font=\scriptsize\color{black}, anchor=center]
        at (1.35, 0.18) {(a) Bidding};

    \node[assigned]   (v1) at (0.00,  0.00) {$v_1$};
    \node[assigned]   (v2) at (0.40, -0.38) {$v_2$};
    \node[unassigned] (v3) at (0.00, -0.76) {$v_3$};
    \node[unassigned] (v4) at (0.40, -1.14) {$v_4$};
    \node[unassigned] (v5) at (0.00, -1.52) {$v_5$};

    \node[cluster] (c1) at (2.20, -0.38) {};
    \node[font=\tiny\bfseries] at (2.28, -0.24) {$C_1$};
    \node[pricelabel]          at (2.28, -0.48) {\scalebox{0.8}{$p_1$}};
    \node[slotfilled] at (1.88, -0.28) {};
    \node[slotempty]  at (1.88, -0.48) {};

    \node[cluster] (c2) at (2.20, -1.14) {};
    \node[font=\tiny\bfseries] at (2.28, -1.00) {$C_2$};
    \node[pricelabel]          at (2.28, -1.28) {\scalebox{0.8}{$p_2$}};
    \node[slotfilled] at (1.88, -0.98) {};
    \node[slotempty]  at (1.88, -1.18) {};
    \node[slotempty]  at (1.88, -1.38) {};

    \begin{pgfonlayer}{background}
        \draw[assignline] (v1) -- (c1);
        \draw[assignline] (v2) -- (c2);
    \end{pgfonlayer}

    \draw[bidarrow] (v3.east) to node[above, font=\tiny] {$b_3$} (c1.west);
    \draw[bidarrow] (v4.east) to node[above, pos = 0.15, font=\tiny] {$b_4$} (c1.190);
    \draw[bidarrow] (v5.east) to node[above, font=\tiny] {$b_5$} (c2.210);

\end{scope}

\begin{scope}[xshift=3.35cm, local bounding box=panel2]

    \node[font=\scriptsize\color{black}, anchor=center]
        at (1.35, 0.18) {(b) Assignment};

    \node[assigned]    (v1b) at (0.00,  0.00) {$v_1$};
    \node[assigned]    (v2b) at (0.40, -0.38) {$v_2$};
    \node[newassigned] (v3b) at (0.00, -0.76) {$v_3$};
    \node[loser]       (v4b) at (0.40, -1.14) {$v_4$};
    \node[newassigned] (v5b) at (0.00, -1.52) {$v_5$};

    \node[cluster] (c1b) at (2.20, -0.38) {};
    \node[font=\tiny\bfseries] at (2.38, -0.24) {$C_1$};
    \node[pricelabel]          at (2.38, -0.48) {\scalebox{0.8}{$p_1\!+\!\delta_1$}};
    \node[slotfilled] at (1.88, -0.28) {};
    \node[slotfilled] at (1.88, -0.48) {};

    \node[cluster] (c2b) at (2.20, -1.14) {};
    \node[font=\tiny\bfseries] at (2.38, -1.00) {$C_2$};
    \node[pricelabel]          at (2.38, -1.28) {\scalebox{0.8}{$p_2\!+\!\delta_2$}};
    \node[slotfilled] at (1.88, -0.98) {};
    \node[slotfilled] at (1.88, -1.18) {};
    \node[slotempty]  at (1.88, -1.38) {};

    \begin{pgfonlayer}{background}
        \draw[assignline] (v1b) -- (c1b);
        \draw[assignline] (v2b) -- (c2b);
    \end{pgfonlayer}

    \draw[arrow, green!50!black] (v3b.east) to (c1b.west);
    \draw[arrow, green!50!black] (v5b.east) to (c2b.210);

    \draw[->, thin, dashed, orange!70!black] (v4b.west) -- ++(-0.14,0)
    node[midway, left, font=\tiny\itshape] {rebids};

\end{scope}

\begin{scope}[xshift=6.85cm, yshift=-0.30cm, local bounding box=legendbox]
    \foreach \col/\lbl/\cy in {
        green!20/Assigned/0.00,
        blue!20/New/{-0.26},
        orange!20/Outbid/{-0.52},
        white/Unassigned/{-0.78}
    }{
        \node[circle, draw=gray!40, thin, fill=\col,
              minimum size=0.15cm, inner sep=0pt,
              label={[font=\tiny, label distance=2pt]right:\lbl}] at (0, \cy) {};
    }
\end{scope}
\begin{pgfonlayer}{background}
    \node[draw=gray!65, thin, fill=white,
          fit=(legendbox), inner sep=2.5pt] {};
\end{pgfonlayer}

\end{tikzpicture}
    }
    \caption{Simplified iteration of the auction algorithm for capacitated cluster assignment.
(a)~Unassigned vectors bid on their preferred cluster; bid value is score minus price ($b_{i,c}$\,=\,$S_{i,c} - p_c$).
(b)~The highest bidder wins the slot; outbid vectors rebid in subsequent iterations. Prices increase after each assignment, ensuring convergence.
}
    \label{fig:auction}
\end{figure}

A linear assignment problem seeks a maximal matching between two equally sized sets given a score for each possible pairing, where each element must be assigned to exactly one counterpart.
A capacitated linear assignment problem is an assignment problem with many-to-one assignment but with a limit on the number of elements assigned to the same bucket. 

In our case, we assign $\nindex$ database vectors to a set of clusters $C$. 
Given a score matrix $S \in \mathbb{R}^{\nindex \times |C|}$, and a quota $Q \ge \nindex / |C|$ on cluster sizes, we solve:
\begin{align}
\label{eq:capacitated_linear_assignment}
\max_{h} \quad  \sum_{\ell=1}^{\nindex} S_{\ell,h(\ell)} \quad 
\text{subject to} \quad
 \left| \{ \ell: h(\ell) = c \} \right| \leq Q \qquad \forall c \in C.
\end{align}

In the general case, this class of assignment problems can be solved using the auction algorithm~\citep{bertsekas1988auction}, which has complexity $\mathcal{O}(|C| \nindex ^2 \log (\nindex))$.

Figure~\ref{fig:auction} illustrates the auction mechanism: database vectors iteratively bid for cluster slots, whose prices rise with demand. When a slot reaches capacity, its lowest bidder is displaced and   
must bid elsewhere. This competition  enforces the capacity constraint, and thus the cluster balance, while converging to an assignment that maximizes the objective.                              
The auction is especially well-suited to our setup: it is highly parallelizable on GPU and converges quickly in practice.                                                                                                                                                      
We provide pseudocode and more details in Appendix~\ref{app:pseudocode}. We refer to Bertsekas~\citep{Bertsekas1989AuctionTransportation} for a complete description.

\section{Related work}
\label{sec:related}

\paragraph{Neural Networks for partition-based indexing.}
Throughout this paper, we focus on partition-based indexing.
Traditional partitioning methods from the literature are based on graphs~\citep{gottesbüren2024unleashinggraphpartitioninglargescale}, LSH~\citep{lshandoni2015practical,lshdasgupta2017neural,lshlv2007multi}, quantization~\citep{sivic2003videogoogle,jegou2010product,babenko2014additive,Niu2023ResidualSearch,chen2010approximate}, or trees~\citep{Sproull1991Refinements,10.1145/1060745.1060840}.
A recent line of work leverages neural networks for efficient partitioning~\citep{dong2020learningspacepartitionsnearest,mazaré2025inferencetimesparseattentionasymmetric,fahim2022unsupervisedspacepartitioningnearest,Gupta2022BLISSAB}.
In Neural LSH~\citep{dong2020learningspacepartitionsnearest}, balanced graph partitioning is applied to the database's k-NN graph,
then a neural network is trained on the database to serve as the probing function. But this approach ignores the query distribution. Moreover, the \emph{partition} and the \emph{probing function} are not optimized jointly with a repeated feedback loop, precluding the reach of optimality.

BLISS~\citep{Gupta2022BLISSAB} and USP~\citep{fahim2022unsupervisedspacepartitioningnearest} address this by alternating optimization between the partition and the probing function; however, both are conflated in a single model which restricts their expressivity and adaptation to the query distribution. Moreover, this conflation creates a moving optimization target that hurts the alternating optimization process, because every weight update modifies both components simultaneously. We compare against these methods experimentally in Section~\ref{sec:experiments}.
More recently, SAAP~\citep{mazaré2025inferencetimesparseattentionasymmetric} employed an asymmetric partition-probing design for sparse attention, but this use-case is outside the scope of our work.

\paragraph{Balanced clusters.}
Balanced clusters reduce latency variance and offer significant computational advantages~\citep{tavenard2011balancing}.
Vanilla K-Means is typically fairly balanced; but Balanced K-Means~\citep{balanced_kmeans} enforces perfect balance by solving a combinatorial assignment problem at each step of Lloyd's algorithm, at the cost of scalability.
Multi-codebook variants such as Product Quantization~\citep{imi_babenko} and Additive Quantization~\citep{babenko2014additive,chen2010approximate} tend to produce highly unbalanced clusters, as they represent the joint distribution of two correlated classifiers — however, to our knowledge, no balanced multi-codebook method has been proposed. Graph partitioning is also a common means to achieve balanced clusters,
with Gottesbüren et al.~\citep{gottesbüren2024unleashinggraphpartitioninglargescale} recently applying it at large scale on k-NN graphs. That work also focused on HNSW integration to accelerate search.

\paragraph{Large partitions.}
When the dataset size increases, the optimal number of clusters should increase, empirically following a power law~\citep{douze2024faiss}. 
This increase in cluster count makes searching centroids themselves slower, so they can be indexed with HNSW~\citep{baranchuk2018revisiting}. 
A more radical approach is to decompose the cluster prediction into a Cartesian product, which avoids storing the centroids altogether~\citep{imi_babenko,lample2019large}. 
We adapt these two approaches in \OURS.

\section{Method: Cluster with Auctions}
\label{sec:method}

This section introduces the \OURS space partition learning.
We introduce our common learning objective in \ref{subsubsec:learn_obj}, optimized in 
alternated steps, described in \ref{subsubsec:backprop}.

\subsection{Supervision}
\label{subsec:supervision}

At training time, we use a set of queries with their ground-truth nearest neighbors, $\mathcal{D}_{\text{train}} = (q_i,\mathcal{N}_{k'}(q_i))_{i=1,\dots,\ntrain}$, where $q_i$ follows the query distribution and $k' \neq k$. $k$ is the retrieval target — the number of NNs returned at search time — while $k'$ is the number of NNs used in training for supervision.
The probing function $f_{\theta}(q_i) \in [0, 1]^{|C|}$ outputs a score for each cluster.

For a query $q_i$ and a fixed assignment $h$, we define $p_{h,q_i}$ as the distribution of $q_i$'s $k'$ nearest neighbors over clusters:
\begin{align}
p_{h,q_i}(c)
=
\frac{1}{k'}\sum_{\ell \in \mathcal{N}_{k'}(q_i)} \mathbf{1}[h(\ell) = c].
\label{eq:phqdef}
\end{align}
The goal of the probing function is to identify which clusters contain the most of $q_i$'s nearest neighbors to maximize target \textit{Recall@k}.
If $k'=k$, this is achieved exactly when $f_{\theta}(q_i)$ matches $p_{h,q_i}$, so we use $p_{h,q_i}$ as the supervision target and train $f_\theta$ by minimizing the cross-entropy $\mathrm{CE}(p_{h,q_i}, f_\theta(q_i))$.
This is similar to Neural LSH~\citep{dong2020learningspacepartitionsnearest}, except that we train on queries from the query distribution whereas Neural LSH focuses on database vectors.

We use $k' \geq k$ in practice: it spreads supervision across more clusters, providing a richer signal and reducing overfitting when training data is scarce; we discuss the choice of $k'$ in Appendix~\ref{subsubsec:exp-scaling}.

\subsection{Learning objective}
\label{subsubsec:learn_obj}

We add a balancedness constraint, which yields the following objective:
\begin{equation}
    \min_{\theta, h} \sum_{i=1}^\ntrain
     \mathrm{CE}\bigl(p_{h,q_i},\, f_{\theta}(q_i)\bigr),
     \quad \text{s.t.} \quad
     \left| \{ \ell \mid h(\ell) = c \} \right| \leq Q, \forall c \in C,
    \label{eq:general_opt}
\end{equation}

where $Q$ is a quota on the size of clusters.
This balancedness constraint ensures that the number of distance computations to the cluster content is capped to $\nprobe \times Q$. 
This learning objective has several advantages.
First, the cross-entropy loss aligns the probing function scores with actual nearest neighbor distributions — a more direct proxy for recall than the reconstruction error minimized by K-Means. Second, it is trained on the query distribution, thus handling the OOD case by design.
Third, the partition (represented by $h$) and the probing function $f_{\theta}$ are strictly separate objects optimized jointly through a \textit{unified loss}. Finally, cluster balancing is strictly enforced through the quota $Q$.
To our knowledge, this is the first partitioning method that combines these properties.

\subsection{Alternating optimization}
\label{sec:altern_opt_method}

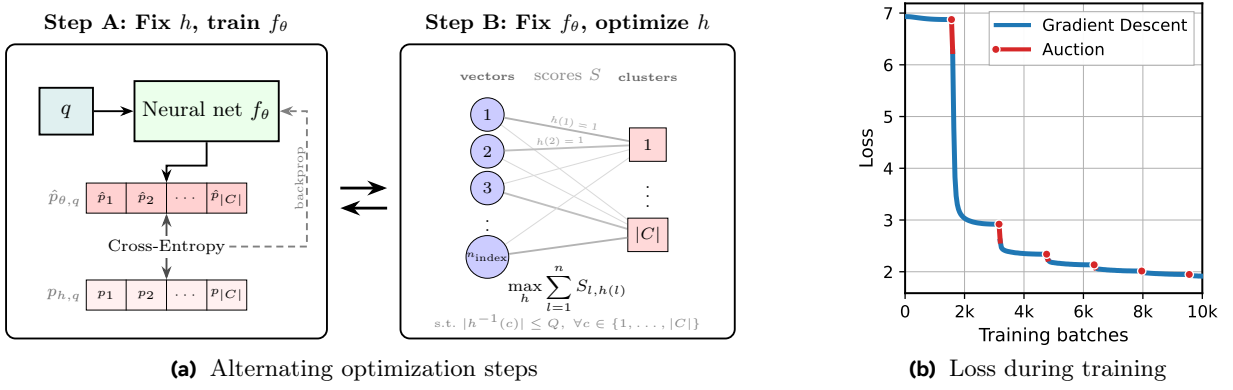
\begin{figure}
\centering
\begin{subfigure}[b]{0.62\textwidth}
    \centering
    \resizebox{\linewidth}{!}{%
        \usetikzlibrary{positioning,shapes.geometric,arrows.meta,calc,matrix}
\begin{tikzpicture}[>=Stealth,
    box/.style={rectangle, draw, thick, align=center},
    arr/.style={-{Stealth[length=2mm]}, thick},
    darr/.style={{Stealth[length=2mm]}-{Stealth[length=2mm]}, thick, black!70}
]

\node[font=\small\bfseries, anchor=south] at (2.0, 2.35) {Step A: Fix $h$, train $f_\theta$};
\draw[thick, rounded corners=4pt] (-0.4, -2.1) rectangle (4.4, 2.3);

\node[box, fill=teal!12, minimum width=0.8cm, minimum height=0.7cm, font=\small] (q) at (0.5, 1.3) {$q$};
\node[box, fill=green!8, minimum width=2.0cm, minimum height=0.9cm, font=\small] (nn) at (2.6, 1.3) {\strut Neural net $f_{\theta}$};
\draw[arr] (q) -- (nn);

\node[rectangle, draw, fill=red!18, minimum width=2.4cm, minimum height=0.45cm, inner sep=0pt] (phat) at (2.0, 0.0) {};
\draw ($(phat.north west)!0.25!(phat.north east)$) -- ($(phat.south west)!0.25!(phat.south east)$);
\draw ($(phat.north west)!0.50!(phat.north east)$) -- ($(phat.south west)!0.50!(phat.south east)$);
\draw ($(phat.north west)!0.75!(phat.north east)$) -- ($(phat.south west)!0.75!(phat.south east)$);
\node[font=\tiny] at ($(phat.west)!0.125!(phat.east)$) {$\hat{p}_1$};
\node[font=\tiny] at ($(phat.west)!0.375!(phat.east)$) {$\hat{p}_2$};
\node[font=\tiny] at ($(phat.west)!0.625!(phat.east)$) {$\cdots$};
\node[font=\tiny] at ($(phat.west)!0.875!(phat.east)$) {$\hat{p}_{|C|}$};
\node[font=\scriptsize, text=black!55, anchor=east] at (0.85, 0.0) {$\hat{p}_{\theta,q}$};

\node[rectangle, draw, fill=red!6, minimum width=2.4cm, minimum height=0.45cm, inner sep=0pt] (psoft) at (2.0, -1.45) {};
\draw ($(psoft.north west)!0.25!(psoft.north east)$) -- ($(psoft.south west)!0.25!(psoft.south east)$);
\draw ($(psoft.north west)!0.50!(psoft.north east)$) -- ($(psoft.south west)!0.50!(psoft.south east)$);
\draw ($(psoft.north west)!0.75!(psoft.north east)$) -- ($(psoft.south west)!0.75!(psoft.south east)$);
\node[font=\tiny] at ($(psoft.west)!0.125!(psoft.east)$) {$p_1$};
\node[font=\tiny] at ($(psoft.west)!0.375!(psoft.east)$) {$p_2$};
\node[font=\tiny] at ($(psoft.west)!0.625!(psoft.east)$) {$\cdots$};
\node[font=\tiny] at ($(psoft.west)!0.875!(psoft.east)$) {$p_{|C|}$};
\node[font=\scriptsize, text=black!55, anchor=east] at (0.85, -1.45) {$p_{h,q}$};

\draw[arr] (nn.south) -- ++(0,-0.35) -| (phat.north);

\draw[darr] (phat) -- (psoft);
\node[font=\scriptsize, fill=white, inner sep=1.5pt](kldiv) at (2.0, -0.725) {Cross-Entropy};

\draw[arr, black!50, densely dashed]
    (kldiv.east) -| (4.1, 1.3) -- (nn.east);
\node[font=\tiny, text=black!50, rotate=90, anchor=south] at (4.2, 0.3) {backprop};

\draw[->, line width=1.2pt, >=Stealth] (4.6, 0.15) -- (5.3, 0.15);
\draw[->, line width=1.2pt, >=Stealth] (5.3, -0.15) -- (4.6, -0.15);

\node[font=\small\bfseries, anchor=south] at (8.0, 2.35) {Step B: Fix $f_\theta$, optimize $h$};
\draw[thick, rounded corners=4pt] (5.5, -2.1) rectangle (10.5, 2.3);

\pgfdeclarelayer{bg}
\pgfsetlayers{bg,main}

\def\bx{8.0}

\node[circle, draw, fill=blue!20, minimum size=0.50cm, inner sep=0pt, font=\scriptsize] (v1) at (\bx-1.2, 1.25) {$1$};
\node[circle, draw, fill=blue!20, minimum size=0.50cm, inner sep=0pt, font=\scriptsize] (v2) at (\bx-1.2, 0.70) {$2$};
\node[circle, draw, fill=blue!20, minimum size=0.50cm, inner sep=0pt, font=\scriptsize] (v3) at (\bx-1.2, 0.15) {$3$};
\node[font=\scriptsize] at (\bx-1.2, -0.38) {$\vdots$};
\node[circle, draw, fill=blue!20, minimum size=0.50cm, inner sep=0pt, font=\tiny] (vn) at (\bx-1.2, -0.88) {\scalebox{0.85}{$\nindex$}};
\node[font=\tiny\bfseries, text=black!60] at (\bx-1.2, 1.82) {vectors};

\node[rectangle, draw, fill=red!15, minimum width=0.55cm, minimum height=0.5cm, inner sep=0pt, font=\scriptsize] (c1) at (\bx+1.2, 0.80) {$1$};
\node[font=\scriptsize] at (\bx+1.2, 0.13) {$\vdots$};
\node[rectangle, draw, fill=red!15, minimum width=0.6cm, minimum height=0.5cm, inner sep=0pt, font=\scriptsize] (ck) at (\bx+1.2, -0.55) {$|C|$};
\node[font=\tiny\bfseries, text=black!60] at (\bx+1.2, 1.82) {clusters};

\begin{pgfonlayer}{bg}
    \draw[-, black!15] (v1) -- (ck);
    \draw[-, black!15] (v2) -- (ck);
    \draw[-, black!15] (v3) -- (c1);
    \draw[-, black!25, thick] (v3) -- (ck);
    \draw[-, black!15] (vn) -- (c1);
\end{pgfonlayer}
\draw[-, black!30, thick] (v1) -- (c1)
    node[font=\tiny, text=black!40, above=0pt, inner sep=0.5pt, pos=0.55, sloped] {\scalebox{0.8}{$h(1)=1$}};
\draw[-, black!25, thick] (v2) -- (c1)
    node[font=\tiny, text=black!40, above=0pt, inner sep=0.5pt, pos=0.45, sloped] {\scalebox{0.8}{$h(2)=1$}};
\draw[-, black!30, thick] (vn) -- (ck);

\node[font=\scriptsize, text=black!45] at (\bx, 1.85) {scores $S$};

\node[font=\scriptsize] at (\bx, -1.35) {$\displaystyle\max_{h} \sum_{l=1}^{n} S_{l,h(l)}$};
\node[font=\tiny, text=black!50] at (\bx, -1.85) {s.t.\ $|h^{-1}(c)| \leq Q,\; \forall c \in \{1,\dots,\vert C \vert\}$};

\end{tikzpicture}%
    }
    \caption{Alternating optimization steps}
\end{subfigure}
\hfill
\begin{subfigure}[b]{0.29\textwidth}
    \centering
    \vspace{6pt}
    \includegraphics[width=\linewidth]{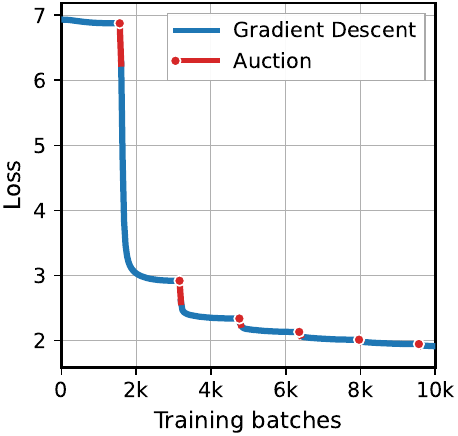}
    \vspace{-15pt}
    \caption{Loss during training}
\end{subfigure}
\caption{Overview of \OURS's alternating optimization.
(a)~The two alternated steps: backpropagation on $f_\theta$ (left) and combinatorial optimization of $h$ via the auction algorithm (right), repeated until convergence.
(b)~Evolution of the evaluation loss during training, starting from a random partition. Red segments mark repartitioning steps via the auction algorithm; the loss decreases monotonically.
}
\label{fig:alternating-optimization}
\end{figure}

This alternating optimization is illustrated in Figure~\ref{fig:alternating-optimization} and detailed in Appendix~\ref{app:altern_opt_detailed}.

\paragraph{Model training via back-propagation.}
\label{subsubsec:backprop}
For a fixed $h$, we can ignore the quota constraint and the objective in Equation~\ref{eq:general_opt} reduces to
\begin{align}
    &\min_{\theta} \sum_{i=1}^{\ntrain} \mathrm{CE}\bigl(p_{h,q_i},\, f_{\theta}(q_i)\bigr).
\end{align}
Because $p_{h,q_i}$ is fixed, this is a classical cross-entropy minimization objective that is solved by back-propagation on the neural network parameters $\theta$.

\paragraph{Optimizing the database assignment.}
\label{subsubsec:iter_repart}

When $\theta$ is fixed, we denote $\hat{p}_{\theta,q_i} = f_{\theta}(q_i)$.
The loss of Equation~\ref{eq:general_opt} decomposes as:
\begin{align}
\sum_{i=1}^{\ntrain} \mathrm{CE}\bigl(p_{h,q_i},\, \hat{p}_{\theta,q_i}\bigr)
&= \sum_{i=1}^{\ntrain} \sum_{c \in C}
  -\log\bigl(\hat{p}_{\theta,q_i}(c)\bigr)\, p_{h,q_i}(c) \\
&= \sum_{i=1}^{\ntrain} \sum_{c \in C}
  -\log\bigl(\hat{p}_{\theta,q_i}(c)\bigr)\,
  \frac{1}{k'}\sum_{\ell=1}^{\nindex}
  \mathbf{1}[\ell \in \mathcal{N}_{k'}(q_i)]\,
  \mathbf{1}[h(\ell) = c] \\
&= \sum_{\ell=1}^{\nindex}
  \underbrace{
    -\frac{1}{k'}\sum_{i=1}^{\ntrain}
    \log\bigl(\hat{p}_{\theta,q_i}(h(\ell))\bigr)\,
    \mathbf{1}[\ell \in \mathcal{N}_{k'}(q_i)]
  }_{\textstyle S_{\ell,\,h(\ell)}}.
\label{eq:score_matrix}
\end{align}

In combination with the quota constraint of Equation~\ref{eq:general_opt},
this is a capacitated linear assignment problem (Equation~\ref{eq:capacitated_linear_assignment}) from $\{1,\dots,\nindex\}$ to $C$ with $S$ as the score matrix. This linear structure is advantageous, because we can efficiently solve it with the auction algorithm (see \ref{subsec:lin_ass_prob}).
Note that computing the matrix $S$ requires a forward pass on the whole training dataset (Equation~\ref{eq:score_matrix}) with an iterative update of the rows of $S$. As we detail in Appendix \ref{app:trick_s}, we use a trick to avoid log and softmax operations, considerably speeding up this step.

The size of $S$ can also become a memory bottleneck for large indexes. 
In practice, we use a preliminary clustering step on the database vectors to restrict which assignments are allowed. 
The effect is to make $S$ sparse, yielding a 5-10$\times$ reduction in memory usage. 
See Appendix~\ref{subsec:sparsification} for details.

\subsection{Architectures for query assignment}
\label{subsec:architectures}

We detail variants of the architecture of $f_\theta: q\in \mathbb{R}^d \mapsto [0,1]^{|C|}$ that we study: 

\paragraph{\FLAT.}

\begin{figure}
    \includegraphics[width=0.99\textwidth, clip, trim={0 7.5cm 0 0}]{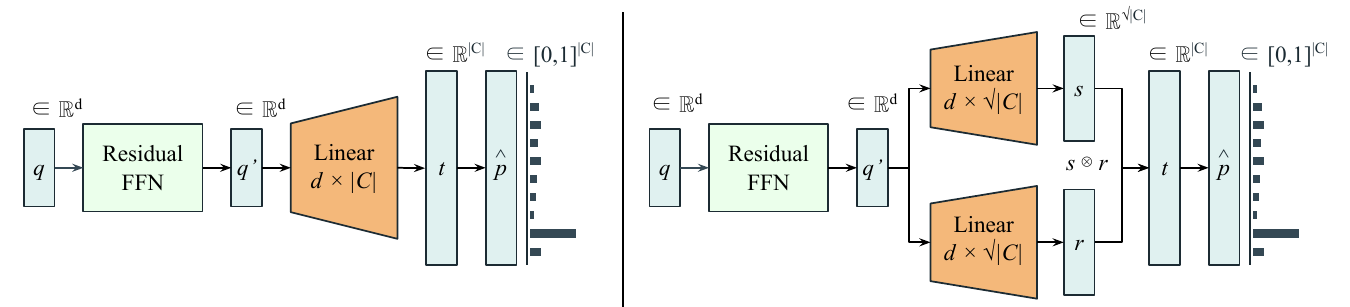}
\caption{
Architectures for the probing function $f_\theta$.
\emph{Left:} \FLAT uses a residual FFN and a linear layer $d \times |C|$ to produce logits $t \in \mathbb{R}^{|C|}$. 
\emph{Right:} \PROD replaces it with two linear layers $d \times \sqrt{|C|}$, yielding $s, r$ of dimension $\sqrt{|C|}$. 
The $|C|$ scores are the weighted pairwise sum of $s$ and $r$ (Equation \ref{eq:prod}), reducing classification parameters from $d|C|$ to $2d\sqrt{|C|}$. 
Softmax is applied to obtain $\hat{p} \in [0,1]^{|C|}$.}
\label{fig:arch}
\end{figure}

The vanilla \FLAT architecture is a residual feed-forward network of $M$ layers followed by a linear layer of output size $|C|$ and a softmax (Figure~\ref{fig:arch}, left).
The FFN layers are gated SiLU with an internal activation of size $2d$.
\FLAT is our analog to the Inverted File Index (IVF)~\citep{sivic2003videogoogle}, with the final linear layer playing the role of centroids.

\paragraph{\HNSW.}

For a large number of IVF centroids ($|C| > 10^4$), identifying the top-$\nprobe$ nearest centroids becomes a bottleneck. 
Then, it is interesting to replace the exhaustive search over centroids with a graph-based index like HNSW~\citep{malkov2018hnsw}. We detail how to combine the graph index with \OURS in Appendix~\ref{app:hnsw_plugged}. 
This speeds up the combination of the last classification layer and top-$\nprobe$ selection: we call it \HNSW.

\paragraph{\PROD.}

To scale to many clusters while keeping a low parameter count, we introduce a \textit{product} architecture (Figure \ref{fig:arch}, right) that forms a large codebook via a Cartesian product of two smaller ones.
This is inspired by the inverted multi-index~\citep{imi_babenko} that uses a product quantizer as a partitioning and probing function.
Rather than one final linear layer of size $d \times |C|$, we pass the final activation $q'\in\mathbb{R}^{d}$ through two linear layers of size $d \times \sqrt{|C|}$.
The two resulting vectors $r\in \mathbb{R}^{\sqrt{|C|}}$ and $s\in \mathbb{R}^{\sqrt{|C|}}$ are then combined using a pairwise transformation $g_\theta(r,s)$ to obtain $|C|$ cluster scores.
Any differentiable function can be used for $g$; we use a pairwise sum weighted by a trainable interaction coefficient matrix $\gamma\in \mathbb{R}^{\sqrt{|C|}\times \sqrt{|C|}}$:
\begin{equation}
g_\theta(r, s)_{i, j} = \gamma_{i,j} \times (r_i + s_j). 
\label{eq:prod}
\end{equation}
This product architecture offers several advantages. 
First, the parameter count is significantly lower because the size of the model is $\mathcal{O}(d\sqrt{|C|} + |C|) \ll \mathcal{O}(d|C|)$.
Second, at training time, the auction algorithm no longer requires to store an $\nindex \times |C|$ score matrix. 
Instead, we only need to store two matrices of size $\nindex \times \sqrt{|C|}$, as we explain in Appendix \ref{app:tricks_prod}.

\section{Experiments}
\label{sec:experiments}
In this section, we evaluate \OURS at three scales: the vanilla \OURS on 1M and 10M vector databases, the graph-indexed \HNSW on 10M, and the product-key \PROD on 100M.

\subsection{Datasets}
\label{subsec:datasets}

We use four benchmark datasets in both in-distribution (ID) and out-of-distribution (OOD) contexts. 
For ID queries, we choose SIFT ($d = 128$, $\ell_2$)~\citep{jegou2010product} and Deep ($d = 96$, $\ell_2$)~\citep{babenko2016efficient}.
To evaluate in an OOD context, we choose the datasets Text-to-Image ($d = 200$, inner-product)~\citep{babenko2021texttoimagedataset} and LAION ($d = 512$, cosine)~\citep{laion_dataset} which exhibit a high query-database distribution discrepancy~\citep{Chen_2024_roargraph}. 
These datasets all provide large sets of training queries, of which we extract subsets for training. 
We use provided evaluation sets of 10k queries for SIFT, Deep and Text-to-Image datasets. 
For LAION, 10k queries are sampled from the original dataset.

\textbf{Metrics.}
We measure the accuracy using Recall@10. 
The search efficiency is reported as either 
\emph{selectivity}, the fraction of the dataset that has been visited (hardware independent), or
directly \emph{QPS}, the number of queries per second on a reference machine. 
We adjust the tradeoff between recall and efficiency by varying the number $\nprobe$ of top-scored clusters to visit per query.
This yields smooth curves (all reported in Appendix \ref{app:add_results}), which we summarize by fixing the recall and reporting the selectivity or QPS.

\subsection{Experimental details}
\label{subsec:exp-details}

\textbf{Training.}
We apply $A$=20 steps of alternated back-propagation (1600 training batches of size 4096) and auction (10k bidding steps). 
A final backpropagation step is performed over 80k batches. 
We train on $\ntrain$=5M queries (resp. 20M) for \FLAT experiments at scale $\nindex$=1M (resp. 10M). 
We train \PROD at scale $\nindex$=100M, on $\ntrain$=50M. 
The supervision width is set at $k'=50$ (resp.\ $100$) for \FLAT at scale 1M (resp.\ 10M) and $k'=100$ for \PROD. The values of $k'$ and $\ntrain$ were chosen following the scaling law provided in Appendix~\ref{subsubsec:exp-scaling}: it shows that training data requirements grow significantly slower than index size — scaling the index by $100\times$ requires only $4\times$ more training queries — making \OURS practical even at large scale.

\textbf{Optimization.} In all experiments, the Adam optimizer starts from learning rate $3 \times 10^{-3}$ and follows a cosine schedule. 
We perform training on NVIDIA-H100 GPUs with 80GB RAM: 
one GPU for $\nindex$=1M, 8 GPUs for $\nindex$=10M and $\nindex$=100M.

\textbf{Initialization of $h$.} 
The number of iterations to convergence depends on the initialization of the partition. 
Therefore, to accelerate convergence, $h$ is initialized from a traditional K-Means in \FLAT, while for \PROD, $h$ starts from product quantization assignments. 
This means that the optimization starts from the K-Means IVF or IMI baseline respectively. 

\textbf{Timings.}
Throughput is measured on an Intel Xeon 6342 CPU at 2.80GHz, in a single thread to minimize variance.
The query batch size is set at $128$. 
Training and inference are implemented in PyTorch, except the top-$k$ computation in the \PROD, which is hand optimized in C++ (see Appendix~\ref{app:avxoptim}). 
To actually compute the search results, we feed the cluster predictions to the search\_preassigned function in Faiss.

\begin{table}
\centering
\caption{QPS at fixed Recall@10$=$0.8 on 1M-vector datasets, for $\vert C\vert \in \{256, 1024\}$ clusters. Higher is better.}
\smallskip
\label{tab:qps_1m}
\resizebox{\textwidth}{!}{
\begin{tabular}{@{\ \ }l rr rr rr rr @{\ \ }}
\toprule
  & \multicolumn{2}{c}{\textbf{Deep1M}} & \multicolumn{2}{c}{\textbf{SIFT1M}} & \multicolumn{2}{c}{\textbf{TextToImage1M}} & \multicolumn{2}{c}{\textbf{LAION1M}} \\
  \cmidrule(lr){2-3} \cmidrule(lr){4-5} \cmidrule(lr){6-7} \cmidrule(lr){8-9}
  \textbf{Method} & $\vert C\vert{=}256$ & $\vert C\vert{=}1024$ & $\vert C\vert{=}256$ & $\vert C\vert{=}1024$ & $\vert C\vert{=}256$ & $\vert C\vert{=}1024$ & $\vert C\vert{=}256$ & $\vert C\vert{=}1024$ \\
\midrule
  K-Means    & 4309 & 9161 & 2495 & 5274 & 1338 & 3592 & 276  & 811  \\
  USP        & 4487 & 6058 & 2552 & 1637 & 453  & 764  & ---  & ---  \\
  Neural LSH & 4596 & 8279 & 2565 & 4937 & 910  & 1766 & 55   & 87   \\
  BLISS      & 2454 & 3285 & 340  & 886  & 229  & 550  & 36   & 66   \\[2pt]
\rowcolor{ourcol}
  \textbf{CwA}, 0 FFN layers & 5334 & \textbf{11664} & 2783 & \textbf{6151} & 2649 & 6615          & 1225          & \textbf{3210} \\
\rowcolor{ourcol}
  \textbf{CwA}, 2 FFN layers & \textbf{5484} & 11262 & \textbf{2831} & 6118 & \textbf{3240} & \textbf{6881} & \textbf{1295} & 2980 \\
\bottomrule
\end{tabular}
}
\end{table}

\subsection{Experiments on \FLAT}
\label{subsec:exp-flat}

We compare vanilla \OURS with a regular K-Means baseline and state of the art methods USP~\citep{fahim2022unsupervisedspacepartitioningnearest} and Neural LSH~\citep{dong2020learningspacepartitionsnearest}, as well as BLISS~\citep{Gupta2022BLISSAB}. 
We use two variants of \OURS: 
one with $M$=0 residual FFN blocks, \ie a linear layer with as many parameters as K-Means; 
one with $M$=2 FFN blocks, \ie a deep neural network comparable to USP, Neural LSH or BLISS. We provide implementation details for the baselines in Appendix~\ref{app:baselines_details}.

Increasing $M$ yields better selectivity-recall tradeoffs, at the cost of higher inference cost. 
Increasing the model's depth beyond $M=2$ does not improve the speed-accuracy tradeoff. 
This is consistent with other works that model a high dimensional vector distribution with neural nets for vector search~\citep{sablayrolles2018spreading,Morozov2019UnsupervisedSearch}.
See Appendix \ref{app:param_count} for details of the parameter counts and forward pass times.

\paragraph{256 and 1024 clusters.}

We first include experiments with $\vert C\vert$=256 and $\vert C\vert$=1024 to match the setting of state-of-the-art methods that cannot scale beyond these values.

\OURS consistently achieves the best throughput across all settings (Table~\ref{tab:qps_1m}, full curves in Appendix~\ref{app:add_results}).
On ID datasets, \OURS outperforms all baselines, achieving up to $1.3\times$ the throughput of the best baseline on Deep. Notably, \OURS (0 FFN layers) --- a linear model --- outperforms Neural LSH and USP. It achieves higher throughput and lower selectivity despite their higher parameter count (Appendix~\ref{app:add_results} and ~\ref{app:param_count}).
The gap widens on OOD datasets: neural baselines collapse below K-Means, while \OURS achieves $2\times$ and $4.7\times$ the throughput of K-Means on Text-to-Image and LAION, respectively.
Consistent with prior findings~\citep{gottesbüren2024unleashinggraphpartitioninglargescale}, BLISS underperforms across all settings, likely due to its heuristic partitioning scheme. 
We note that BLISS, like LSH approaches, is designed to run with multiple partitions simultaneously, which we do not evaluate here.
Training times of \OURS are comparable to other neural baselines (Appendix~\ref{app:training_times}) but remain one or two orders of magnitude higher than K-Means.

\paragraph{Coupling with a graph index.}
\label{subsubsec:exp-hnsw}

We scale to $\nindex$=10M and $|C|$=65536 clusters \footnote{Only the baselines IVF and BLISS scaled to this size; however, we did not include BLISS in further experiments due to its poor performance.}.
In this setting, the computation of $f_\theta$ dominates the cost of distance computations with database vectors (see Table \ref{tab:breakdown_recall2} in the Appendix), which is why we use \HNSW. 
We plug the HNSW on \OURS with $M$=0 FFN layers, to provide a fair comparison to IVF-HNSW which uses the same number of parameters. 

Table~\ref{tab:qps_hnsw} shows that IVF-HNSW improves throughput compared to traditional IVF by accelerating the centroid search, and \HNSW further improves it by combining faster centroid search with better underlying partitions. 

\begin{table}
  \centering
  \caption{QPS at fixed Recall@10$=$0.8 on 10M-vector datasets with $\vert C\vert{=}65536$ clusters. Higher is better.}
  \smallskip
  \label{tab:qps_hnsw}
  {\small
  \begin{tabular}{@{\ \ }l rrr @{\ \ }}
  \toprule
    \textbf{Method} & \textbf{Deep10M} & \textbf{TextToImage10M} & \textbf{LAION10M} \\
  \midrule
    IVF (K-Means) & 2514           & 1667           & 677  \\
    IVF-HNSW      & 6271           & 2662           & 723  \\
  \rowcolor{ourcol} \textbf{CwA} & 2762 & 2186 & 1147 \\
  \rowcolor{ourcol} \textbf{CwA-HNSW} & \textbf{9954} & \textbf{6210} & \textbf{3419} \\
  \bottomrule
  \end{tabular}}
  \end{table}

\begin{table}
\centering
\caption{
Performance of partitioning functions with $\mathcal{O}(\sqrt{|C|})$ centroids. 
We fix $\vert C\vert=512^2$ centroids, recall $\in \{0.7, 0.9\}$, and report the Queries Per Second (QPS). }
\smallskip
\label{tab:fixed_recall}
\small
\setlength{\tabcolsep}{4pt}
\begin{tabular}{l cccc cccc}
\toprule
& \multicolumn{2}{c}{\textbf{Deep100M}}
& \multicolumn{2}{c}{\textbf{SIFT100M}}
& \multicolumn{2}{c}{\textbf{TextToImage100M}}
& \multicolumn{2}{c}{\textbf{LAION100M}} \\
\cmidrule(lr){2-3} \cmidrule(lr){4-5} \cmidrule(lr){6-7} \cmidrule(lr){8-9}
\textbf{Method} 
& R=0.7 & R=0.9
& R=0.7 & R=0.9
& R=0.7 & R=0.9
& R=0.7 & R=0.9 \\
\midrule
RQ
& 1180.7 & 350.8
& 994.4  & 330.0
& 118.3  & 23.3
& ---    & ---   \\
IMI
& 281.3  & 83.0
& 436.7  & 131.5
& 51.6   & 12.7
& 11.4   & ---   \\
\rowcolor{ourcol}
\textbf{\PROD}
& \textbf{2729.6} & \textbf{613.7}
& \textbf{1906.7} & \textbf{446.5}
& \textbf{1671.3} & \textbf{183.3}
& \textbf{1897.7} & \textbf{174.1} \\
\bottomrule
\end{tabular}
\end{table}

\subsection{Experiments on \PROD}
\label{subsec:exp-prod}

We now scale to large databases ($\nindex$=100M) with a high number of clusters ($\vert C\vert = 512^2 \approx$ 262k), where fine-grained partitioning requires \PROD. 
We compare \PROD (with $M$=2 FFN layers) to baselines that also store $\mathcal{O}(\sqrt{|C|})$ centroids.
Thus, we choose as baselines Residual Quantization~\citep{chen2010approximate} and the Inverted Multi-Index~\citep{imi_babenko}. 
We leave out hierarchical variants of K-Means or Neural LSH since they have $\mathcal{O}(|C|)$ stored centroids.

\begin{figure}
    \centering
    \begin{subfigure}[b]{0.32\linewidth}
        \includegraphics[width=\linewidth]{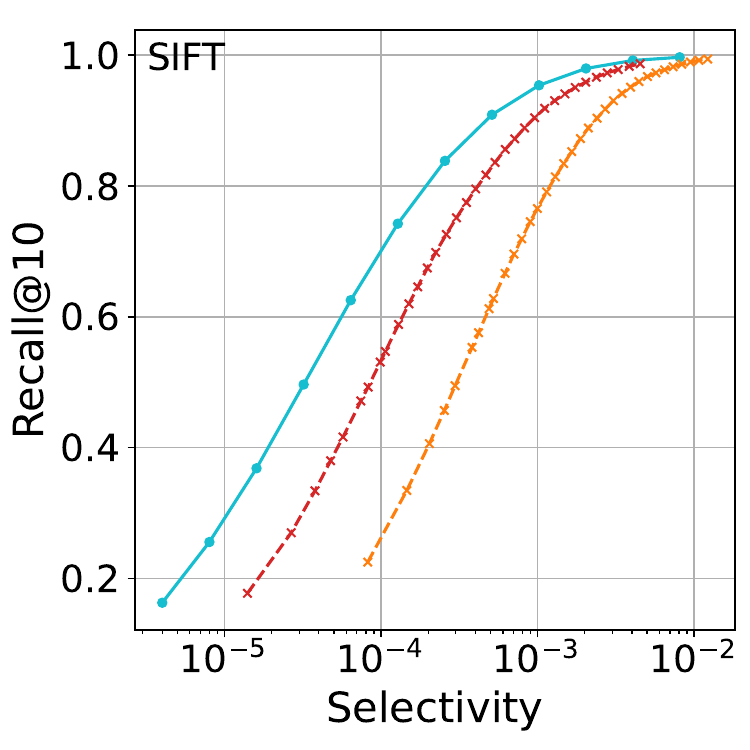}
    \end{subfigure}\hfill
    \begin{subfigure}[b]{0.32\linewidth}
        \includegraphics[width=\linewidth]{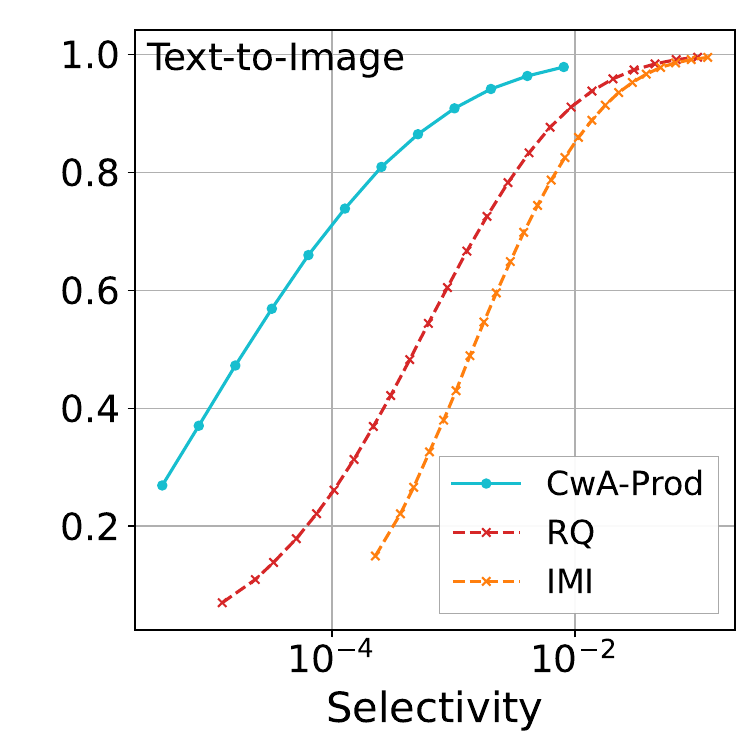}
    \end{subfigure}\hfill
    \begin{subfigure}[b]{0.32\linewidth}
        \includegraphics[width=\linewidth]{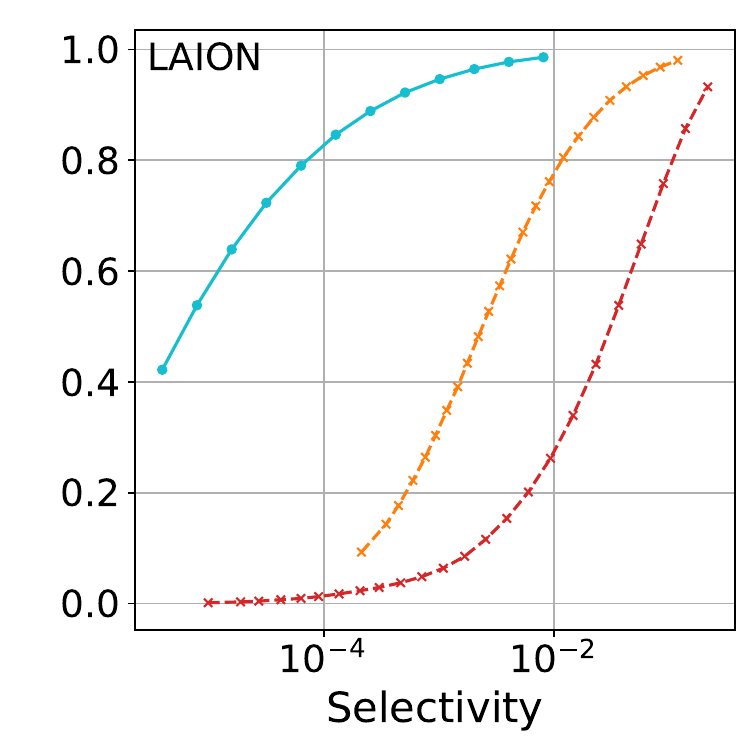}
    \end{subfigure}
    \caption{Performance comparison with baselines on 262k clusters ($n_{\text{index}}$\,$=10^8$): SIFT, Text-to-Image, LAION.}
    \label{fig:sel_262k}
\end{figure}

Table \ref{tab:fixed_recall} breaks down QPS and selectivity values at fixed recall values. 
In the In-Distribution setting (SIFT and Deep), \PROD achieves far better throughput, with up to 2.5$\times$ improvement compared to the best performing baseline. 
In the out-of-distribution setting (Text-to-Image and LAION), the performance of the baselines RQ and IMI is more severely degraded due to the query-database discrepancy. 
The selectivity-recall tradeoff curves in Figure~\ref{fig:sel_262k} show the edge of \PROD on a wide array of operating points.

\begin{figure}[h!]
    \begin{minipage}[t]{0.495\textwidth}
        \includegraphics[width=\linewidth]{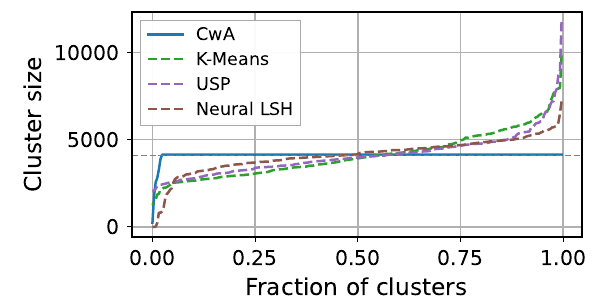}
    \end{minipage}
    \hfill
    \begin{minipage}[t]{0.495\textwidth}
        \includegraphics[width=\linewidth]{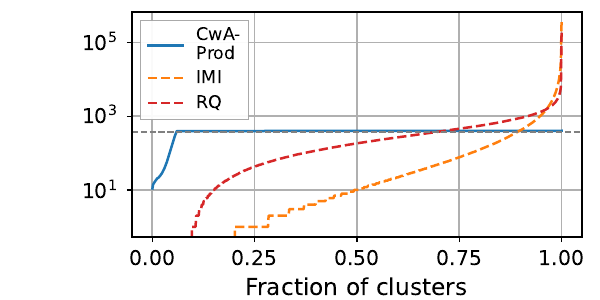}
    \end{minipage}
    \caption{%
        Cluster size distributions on the Deep dataset. \textit{Left:} \OURS ($\nindex=1\text{M}$, $\vert C\vert=256$). \textit{Right:} \PROD ($\nindex=100\text{M}$, $\vert C\vert=512^2=262144$). The grey dashed line marks the uniform distribution; \OURS's distributions remain close to this baseline.
    }
    \label{fig:cluster_dists_deep1B}
\end{figure}

\subsection{Balanced clusters}
\label{subsec:balancing}

\OURS enforces cluster balance through capacitated linear assignment, with a quota on each cluster size. In practice, we use $Q = 1.05 \frac{\nindex}{|C|}$ in all experiments, enforcing near-perfect balance.
Figure \ref{fig:cluster_dists_deep1B} shows the distribution of clusters in the single-codebook (left) and multi-codebook (right) cases: \OURS's clusters are the most uniform in size.
The capping of cluster sizes avoids having some queries with disproportionately high candidate lists, a known limitation of K-Means~\citep{tavenard2011balancing}.

\section{Conclusion}
\label{sec:conclusion}

\OURS is a space partitioning method that outperforms previous works.
\OURS strictly separates the partition from the probing function, allowing each component to specialize to its task. They are optimized in an alternating fashion, with steps of backpropagation and combinatorial assignment.
We applied our framework to the distance metrics of the reference benchmarks (Euclidean distance and inner-product). 
However, our method is agnostic to the metric and can be applied to any sort of relation, provided that ground-truth neighbors are obtainable for the training queries. 

\textbf{Limitations.} This work is limited to evaluating a single partition. 
The combination of several partitions for ANNS, classical in LSH and explored by USP~\citep{fahim2022unsupervisedspacepartitioningnearest} and BLISS~\citep{Gupta2022BLISSAB}, could be an extension of our work. 
The challenge is then to make these partitions complementary.
Finally, we report results only on fixed databases. An interesting extension would be an out-of-sample variant of CWA capable of assigning new database vectors to existing partitions.

\newpage 

\bibliographystyle{assets/plainnat}
\bibliography{refs}

@string{arxiv       = {arXiv preprint}}

@string{iccv      = {ICCV}}

@string{eccv      = {ECCV}}

@string{cvpr      = {CVPR}}

@string{vldb      = {International Conference on Very Large DataBases}}

@string{www       = {WWW: Proceeding of the International Conference on World Wide Web}}

@string{iclr      = {ICLR}}

@STRING{ieeetpami = {{\sc IEEE} Transactions on Pattern Analysis and Machine Intelligence}}

@inproceedings{babenko2016efficient,
  title={Efficient indexing of billion-scale datasets of deep descriptors},
  author={Babenko, Artem and Lempitsky, Victor},
  booktitle=cvpr,
  year={2016}
}

@inproceedings{NEURIPS2018_229754d7_ipnsw,
 author = {Morozov, Stanislav and Babenko, Artem},
 booktitle = {Advances in Neural Information Processing Systems},
 editor = {S. Bengio and H. Wallach and H. Larochelle and K. Grauman and N. Cesa-Bianchi and R. Garnett},
 pages = {},
 publisher = {Curran Associates, Inc.},
 title = {Non-metric Similarity Graphs for Maximum Inner Product Search},
 url = {https://proceedings.neurips.cc/paper_files/paper/2018/file/229754d7799160502a143a72f6789927-Paper.pdf},
 volume = {31},
 year = {2018}
}

@article{kahip,
  author       = {Peter Sanders and
                  Christian Schulz},
  title        = {Think Locally, Act Globally: Perfectly Balanced Graph Partitioning},
  journal      = {CoRR},
  volume       = {abs/1210.0477},
  year         = {2012},
  url          = {http://arxiv.org/abs/1210.0477},
  eprinttype   = {arXiv},
  eprint       = {1210.0477},
  bibsource    = {dblp computer science bibliography, https://dblp.org}
}

@article{jegou2010product,
  title={Product quantization for nearest neighbor search},
  author={J\'egou, Herv\'e and Douze, Matthijs and Schmid, Cordelia},
  journal=ieeetpami,
  volume={33},
  number={1},
  pages={117--128},
  year={2010}
}

@inproceedings{babenko2014additive,
title={Additive quantization for extreme vector compression},
  author={Babenko, Artem and Lempitsky, Victor},
  booktitle=cvpr,
  year={2014}
}

@article{sablayrolles2018spreading,
  title={Spreading vectors for similarity search},
  author={Sablayrolles, Alexandre and Douze, Matthijs and Schmid, Cordelia and J{\'e}gou, Herv{\'e}},
  journal=iclr,
  year={2019}
}

@inproceedings{Morozov2019UnsupervisedSearch,
    title = {Unsupervised Neural Quantization for Compressed-Domain Similarity Search},
    year = {2019},
    booktitle = iccv,
    author = {Morozov, Stanislav and Babenko, Artem}
}

@article{douze2024faiss,
  title={The {Faiss} library},
  author={Douze, Matthijs and Guzhva, Alexandr and Deng, Chengqi and Johnson, Jeff and Szilvasy, Gergely and Mazar{\'e}, Pierre-Emmanuel and Lomeli, Maria and Hosseini, Lucas and J{\'e}gou, Herv{\'e}},
  journal=arxiv, 
volume = {2401.08281},
  year={2024}
}

@inproceedings{pizzi2022self,
  title={A Self-supervised Descriptor for Image Copy Detection},
  author={Pizzi, Ed and Roy, Sreya Dutta and Ravindra, Sugosh Nagavara and Goyal, Priya and Douze, Matthijs},
  booktitle=cvpr,
  year={2022}
}

@article{malkov2018hnsw,
  title={Efficient and Robust Approximate Nearest Neighbor Search Using Hierarchical Navigable Small World Graphs},
  author={Malkov, Yu A. and Yashunin, Dmitry A.},
  journal=ieeetpami,
  volume={42},
  number={4},
  pages={824--836},
  year={2018},
  publisher={IEEE}
}

@inproceedings{baranchuk2018revisiting,
  title={Revisiting the inverted indices for billion-scale approximate nearest neighbors},
  author={Baranchuk, Dmitry and Babenko, Artem and Malkov, Yury},
  booktitle=eccv,
  year={2018}
}

@article{chen2010approximate,
  title={{Approximate nearest neighbor search by residual vector quantization}},
  author={Chen, Yongjian and Guan, Tao and Wang, Cheng},
  journal={Sensors},
  volume={10},
  number={12},
  pages={11259--11273},
  year={2010},
  publisher={Molecular Diversity Preservation International (MDPI)}
}

@article{Niu2023ResidualSearch,
    title = {Residual Vector Product Quantization for Approximate Nearest Neighbor Search},
    year = {2023},
    journal = {Expert Systems with Applications},
    author = {Niu, Lushuai and Xu, Zhi and Zhao, Longyang and He, Daojing and Ji, Jianqiu and Yuan, Xiaoli and Xue, Mian},
    volume = {232}
}

@article{laion_dataset,
  author       = {Christoph Schuhmann and
                  Richard Vencu and
                  Romain Beaumont and
                  Robert Kaczmarczyk and
                  Clayton Mullis and
                  Aarush Katta and
                  Theo Coombes and
                  Jenia Jitsev and
                  Aran Komatsuzaki},
  title        = {{LAION-400M:} Open Dataset of CLIP-Filtered 400 Million Image-Text
                  Pairs},
  journal      = {CoRR},
  volume       = {abs/2111.02114},
  year         = {2021},
  url          = {https://arxiv.org/abs/2111.02114},
  eprinttype    = {arXiv},
  eprint       = {2111.02114},
  bibsource    = {dblp computer science bibliography, https://dblp.org}
}

@misc{babenko2021texttoimagedataset,
  author       = {Baranchuk, Dmitry and Babenko, Artem},
  title        = {Text-to-Image Dataset for Billion-Scale Similarity Search},
  year         = {2021},
  howpublished = {\url{https://research.yandex.com/datasets/text-to-image-dataset-for-billion-scale-similarity-search}},
  note         = {Accessed: August 23, 2023}
}

@article{Chen_2024_roargraph,
   title={RoarGraph: A Projected Bipartite Graph for Efficient Cross-Modal Approximate Nearest Neighbor Search},
   volume={17},
   ISSN={2150-8097},
   url={http://dx.doi.org/10.14778/3681954.3681959},
   DOI={10.14778/3681954.3681959},
   number={11},
   journal={Proceedings of the VLDB Endowment},
   publisher={Association for Computing Machinery (ACM)},
   author={Chen, Meng and Zhang, Kai and He, Zhenying and Jing, Yinan and Wang, X. Sean},
   year={2024},
   month=jul, pages={2735–2749} }

@INPROCEEDINGS{imi_babenko,
  author={Babenko, Artem and Lempitsky, Victor},
  booktitle={2012 IEEE Conference on Computer Vision and Pattern Recognition}, 
  title={The inverted multi-index}, 
  year={2012},
  volume={},
  number={},
  pages={3069-3076},
  doi={10.1109/CVPR.2012.6248038}}

@misc{douze2025machine,
      title={Machine learning and high dimensional vector search}, 
      author={Matthijs Douze},
      year={2025},
      eprint={2502.16931},
      archivePrefix={arXiv},
      primaryClass={cs.LG},
      url={https://arxiv.org/abs/2502.16931}, 
}

@inproceedings{sivic2003videogoogle,
  author    = {Josef Sivic and Andrew Zisserman},
  title     = {Video Google: A text retrieval approach to object matching in videos},
  booktitle = {Proceedings of the International Conference on Computer Vision (ICCV)},
  year      = {2003},
  pages     = {1470--1477}
}

@article{Bertsekas1989AuctionTransportation,
  author    = {Dimitri P. Bertsekas},
  title     = {The Auction Algorithm for the Transportation Problem},
  journal   = {Annals of Operations Research},
  volume    = {20},
  pages     = {67--96},
  year      = {1989},
  month     = {December},
  doi       = {10.1007/BF02021667}
}

@article{bertsekas1988auction,
  author    = {Bertsekas, Dimitri P.},
  title     = {The Auction Algorithm: A Distributed Relaxation Method
               for the Assignment Problem},
  journal   = {Annals of Operations Research},
  volume    = {14},
  number    = {1},
  pages     = {105--123},
  year      = {1988},
  publisher = {Springer},
  doi       = {10.1007/BF02186476}
}

@article{Gupta2022BLISSAB,
  title={BLISS: A Billion scale Index using Iterative Re-partitioning},
  author={Gaurav Gupta and Tharun Medini and Anshumali Shrivastava and Alexander J. Smola},
  journal={Proceedings of the 28th ACM SIGKDD Conference on Knowledge Discovery and Data Mining},
  year={2022},
  url={https://api.semanticscholar.org/CorpusID:251518237}
}

@misc{fahim2022unsupervisedspacepartitioningnearest,
      title={Unsupervised Space Partitioning for Nearest Neighbor Search}, 
      author={Abrar Fahim and Mohammed Eunus Ali and Muhammad Aamir Cheema},
      year={2022},
      eprint={2206.08091},
      archivePrefix={arXiv},
      primaryClass={cs.LG},
      url={https://arxiv.org/abs/2206.08091}, 
}

@misc{dong2020learningspacepartitionsnearest,
      title={Learning Space Partitions for Nearest Neighbor Search}, 
      author={Yihe Dong and Piotr Indyk and Ilya Razenshteyn and Tal Wagner},
      year={2020},
      eprint={1901.08544},
      archivePrefix={arXiv},
      primaryClass={cs.LG},
      url={https://arxiv.org/abs/1901.08544}, 
}

@inproceedings{balanced_kmeans,
author = {Malinen, Mikko},
year = {2014},
month = {08},
pages = {},
title = {Balanced K-Means for Clustering},
booktitle = {Structural, Syntactic, and Statistical Pattern Recognition},
isbn = {978-3-662-44414-6},
doi = {10.1007/978-3-662-44415-3_4}
}

@misc{mazaré2025inferencetimesparseattentionasymmetric,
      title={Inference-time sparse attention with asymmetric indexing}, 
      author={Pierre-Emmanuel Mazaré and Gergely Szilvasy and Maria Lomeli and Francisco Massa and Naila Murray and Hervé Jégou and Matthijs Douze},
      year={2025},
      eprint={2502.08246},
      archivePrefix={arXiv},
      primaryClass={cs.CL},
      url={https://arxiv.org/abs/2502.08246}, 
}

@misc{gottesbüren2024unleashinggraphpartitioninglargescale,
      title={Unleashing Graph Partitioning for Large-Scale Nearest Neighbor Search}, 
      author={Lars Gottesbüren and Laxman Dhulipala and Rajesh Jayaram and Jakub Lacki},
      year={2024},
      eprint={2403.01797},
      archivePrefix={arXiv},
      primaryClass={cs.DS},
      url={https://arxiv.org/abs/2403.01797}, 
}

@inproceedings{lshandoni2015practical,
  author    = {Alexandr Andoni and Piotr Indyk and Thijs Laarhoven and Ilya Razenshteyn and Ludwig Schmidt},
  title     = {Practical and Optimal LSH for Angular Distance},
  booktitle = {Advances in Neural Information Processing Systems},
  year      = {2015},
  pages     = {1225--1233}
}

@inproceedings{lshlv2007multi,
  author    = {Qin Lv and William Josephson and Zhe Wang and Moses Charikar and Kai Li},
  title     = {Multi-Probe LSH: Efficient Indexing for High-Dimensional Similarity Search},
  booktitle = {Proceedings of the 33rd International Conference on Very Large Data Bases},
  year      = {2007},
  pages     = {950--961},
  publisher = {VLDB Endowment}
}

@article{lshdasgupta2017neural,
  author  = {Sanjoy Dasgupta and Charles F. Stevens and Saket Navlakha},
  title   = {A Neural Algorithm for a Fundamental Computing Problem},
  journal = {Science},
  volume  = {358},
  number  = {6364},
  pages   = {793--796},
  year    = {2017}
}

@article{Sproull1991Refinements,
  author    = {Sproull, Robert F.},
  title     = {Refinements to nearest-neighbor searching in k-dimensional trees},
  journal   = {Algorithmica},
  year      = {1991},
  volume    = {6},
  number    = {1-6},
  pages     = {579--589},
  doi       = {10.1007/BF01759061},
  url       = {https://doi.org/10.1007/BF01759061},
  publisher = {Springer},
}

@inproceedings{10.1145/1060745.1060840,
author = {Bawa, Mayank and Condie, Tyson and Ganesan, Prasanna},
title = {LSH forest: self-tuning indexes for similarity search},
year = {2005},
isbn = {1595930469},
publisher = {Association for Computing Machinery},
address = {New York, NY, USA},
url = {https://doi.org/10.1145/1060745.1060840},
doi = {10.1145/1060745.1060840},
booktitle = {Proceedings of the 14th International Conference on World Wide Web},
pages = {651–660},
numpages = {10},
location = {Chiba, Japan},
series = {WWW '05}
}

@article{lewis2020retrieval,
  title={Retrieval-augmented generation for knowledge-intensive nlp tasks},
  author={Lewis, Patrick and Perez, Ethan and Piktus, Aleksandra and Petroni, Fabio and Karpukhin, Vladimir and Goyal, Naman and K{\"u}ttler, Heinrich and Lewis, Mike and Yih, Wen-tau and Rockt{\"a}schel, Tim and others},
  journal={Advances in neural information processing systems},
  volume={33},
  pages={9459--9474},
  year={2020}
}

@inproceedings{tavenard2011balancing,
  title={Balancing clusters to reduce response time variability in large scale image search},
  author={Tavenard, Romain and J{\'e}gou, Herv{\'e} and Amsaleg, Laurent},
  booktitle={2011 9th International Workshop on Content-Based Multimedia Indexing (CBMI)},
  pages={19--24},
  year={2011},
  organization={IEEE}
}

@article{lample2019large,
  title={Large memory layers with product keys},
  author={Lample, Guillaume and Sablayrolles, Alexandre and Ranzato, Marc'Aurelio and Denoyer, Ludovic and J{\'e}gou, Herv{\'e}},
  journal={Advances in Neural Information Processing Systems},
  volume={32},
  year={2019}
}

\appendix
\clearpage

\begin{center}
    \LARGE{Appendices}
\end{center}

\section{Notations}
\label{sec:notation}

Table~\ref{tab:notation_method} summarizes the main notations used in the paper. 

\begin{table}[H]
    \centering
    \begin{tabular}{ll}
        \toprule
        Symbol & Description \\
        \midrule
        $d$ & Vector dimension ($x_i, q \in \mathbb{R}^d$) \\
        $X = (x_i)$ & Dataset of $\nindex$ vectors \\
        ${\nindex}$ & Number of indexed vectors \\
        $q$ & Query vector \\
        $k$ & Number of NNs to retrieve \\
        $k'$ & Number of NNs used in supervision \\
        $C$ & Set of cluster ids \\
        $\nprobe$ & Number of clusters probed at search time \\
        $h$ & Partitioning function, $h:\{1, \dots, \nindex\}\to C$ \\
        $\mathcal{D}_{\text{train}}$ & Training set $(q_i,\mathcal{N}_{k'}(q_i))_{i=1}^{\ntrain}$ \\
        $\ntrain$ & Number of training queries \\
        $f_\theta$ & Parametric probing function \\
        $\mathcal{N}_{k'}(q)$ & Indices of the $k'$-NN of $q$ in $X$ \\
        $Q$ & Quota: maximum number of vectors per cluster \\
        $S \in \mathbb{R}^{\nindex \times |C|}$ & Score matrix for the linear assignment problem \\
        $p_{h,q}$ & Soft supervision label: distribution of $q$'s NNs over clusters under $h$ \\
        $\hat{p}_{\theta,q}$ & Model output: $f_\theta(q)$, predicted distribution over clusters for $q$ \\
        \bottomrule
    \end{tabular}
    \caption{Main notation used in the paper.
    }
    \label{tab:notation_method}
\end{table}

\section{Full selectivity and QPS curves}
\label{app:add_results}
\begin{figure}[H]
    \centering
    \begin{subfigure}[b]{0.24\linewidth}
        \includegraphics[height=3.3cm]{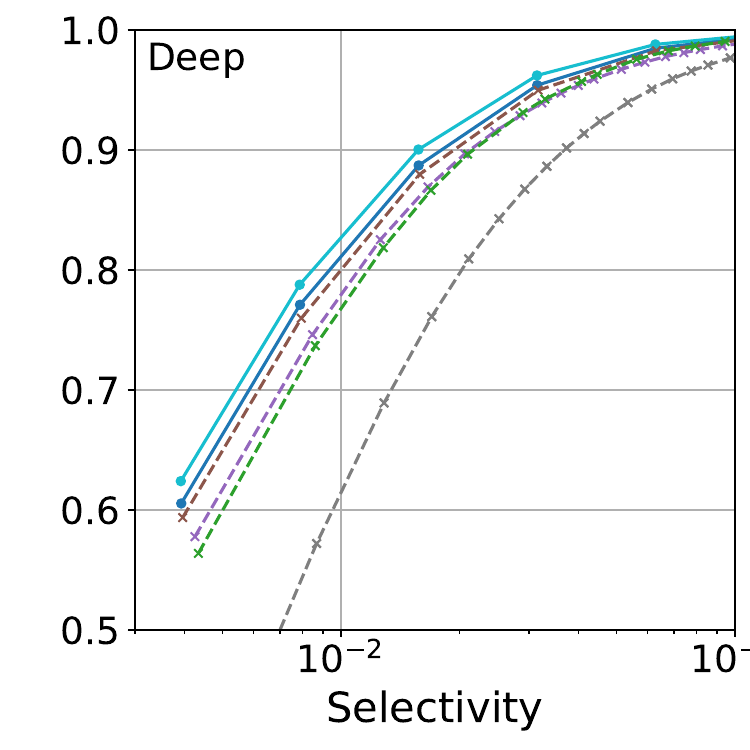}
    \end{subfigure}\hfill
    \begin{subfigure}[b]{0.24\linewidth}
        \includegraphics[height=3.3cm]{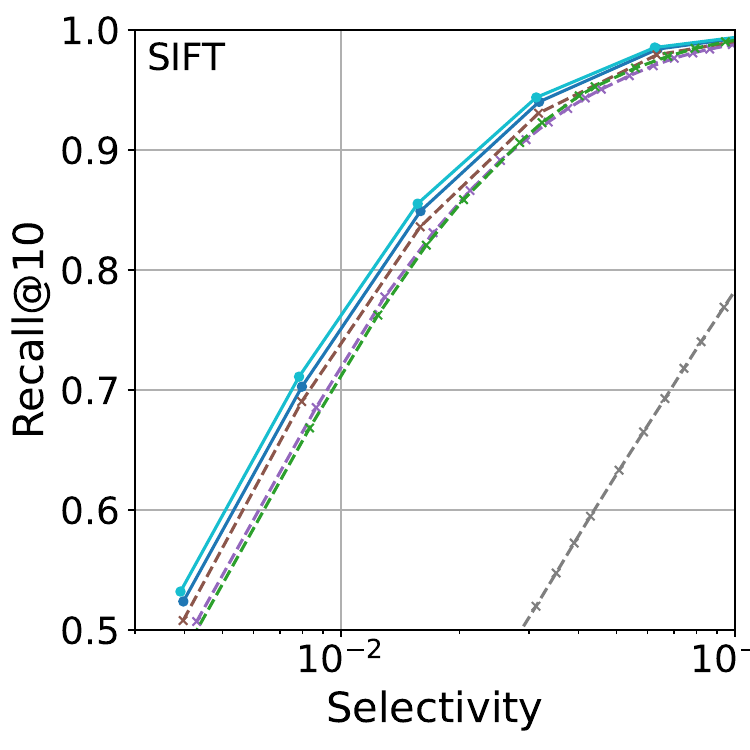}
    \end{subfigure}\hfill
    \begin{subfigure}[b]{0.24\linewidth}
        \includegraphics[height=3.3cm]{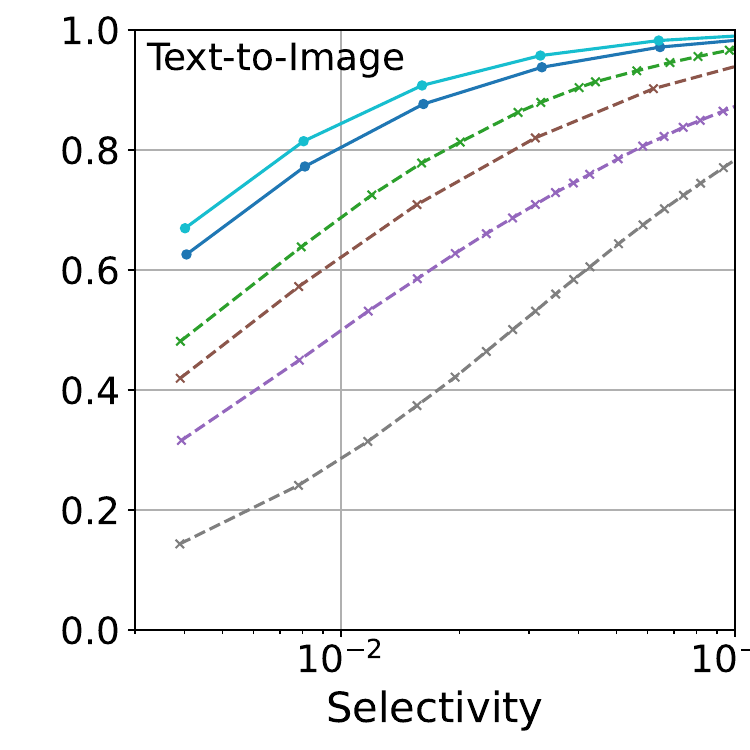}
    \end{subfigure}\hfill
    \begin{subfigure}[b]{0.24\linewidth}
        \includegraphics[height=3.3cm]{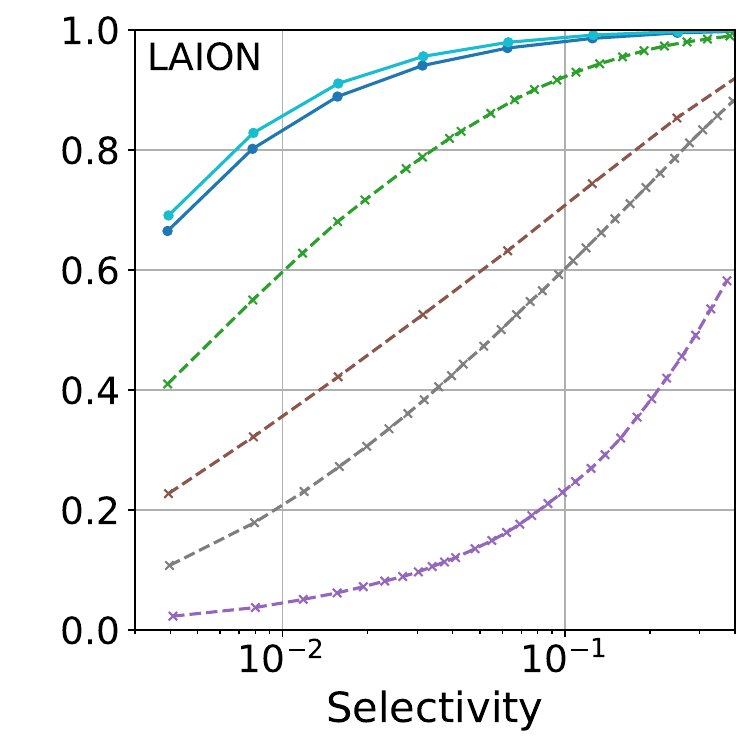}
    \end{subfigure}
    \\[\smallskipamount]
    \includegraphics[height=0.57cm]{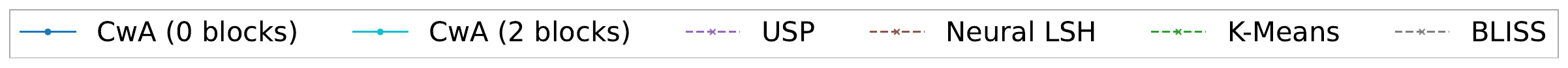}
    \caption{Selectivity vs.\ recall on 256 clusters ($\nindex=1\,048\,576$):
             Deep, SIFT, Text-to-Image, LAION.}
    \label{fig:selectivity_256}
\end{figure}

\begin{figure}[H]
    \centering
    \begin{subfigure}[b]{0.24\linewidth}
        \includegraphics[height=3.3cm]{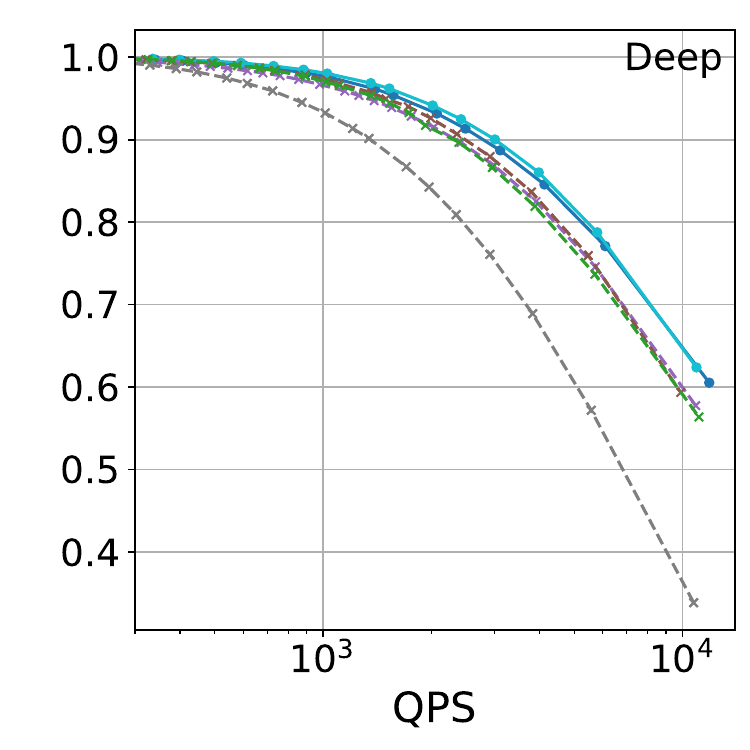}
    \end{subfigure}\hfill
    \begin{subfigure}[b]{0.24\linewidth}
        \includegraphics[height=3.3cm]{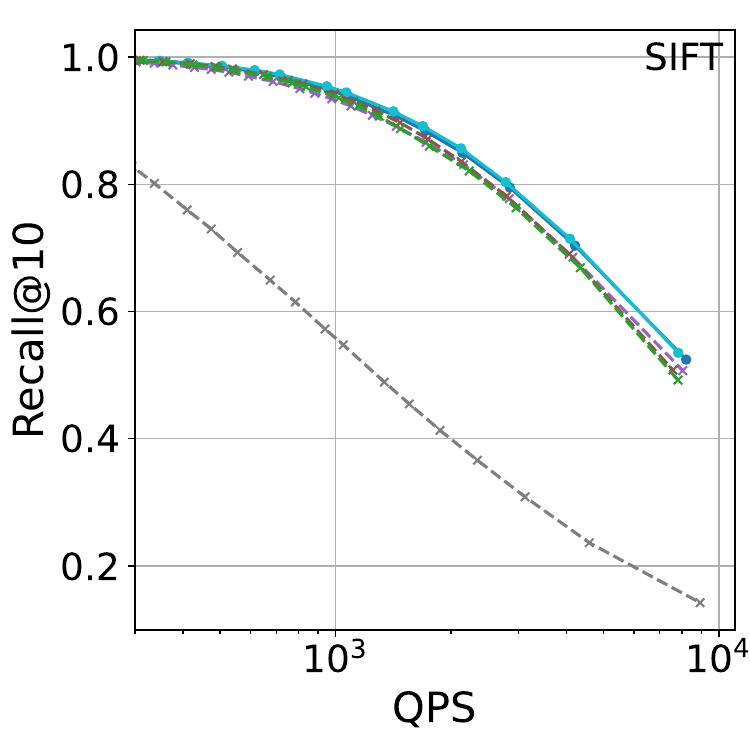}
    \end{subfigure}\hfill
    \begin{subfigure}[b]{0.24\linewidth}
        \includegraphics[height=3.3cm]{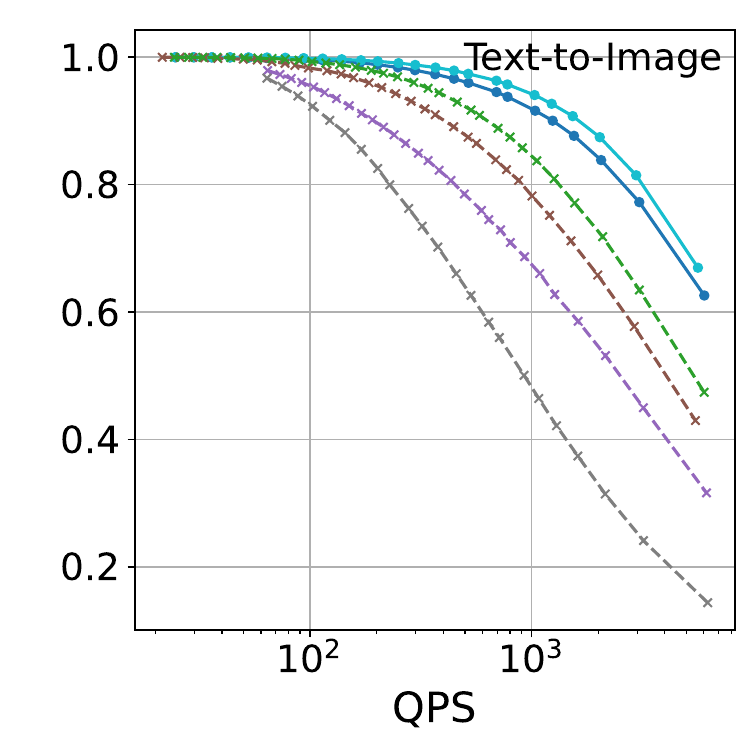}
    \end{subfigure}\hfill
    \begin{subfigure}[b]{0.24\linewidth}
        \includegraphics[height=3.3cm]{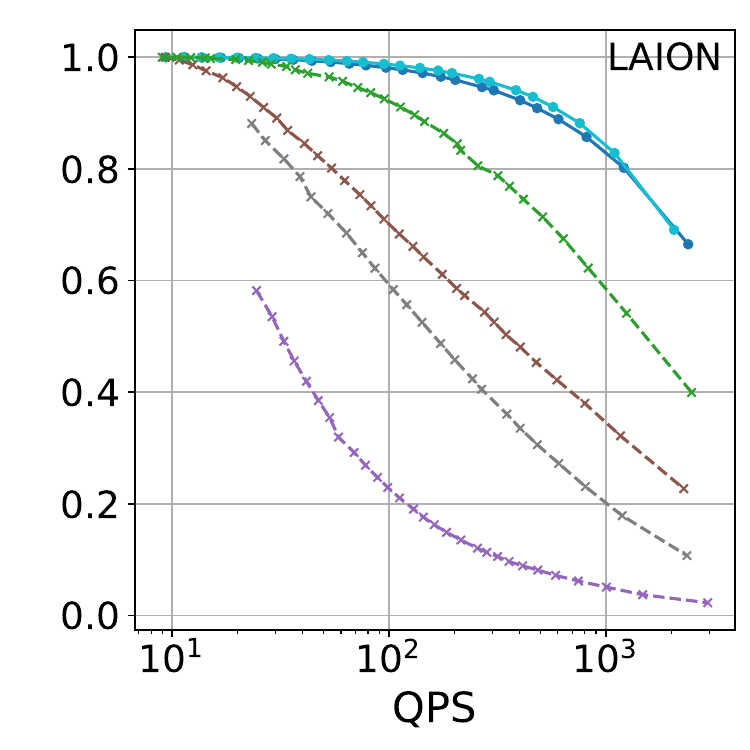}
    \end{subfigure}
    \\[\smallskipamount]
    \includegraphics[height=0.57cm]{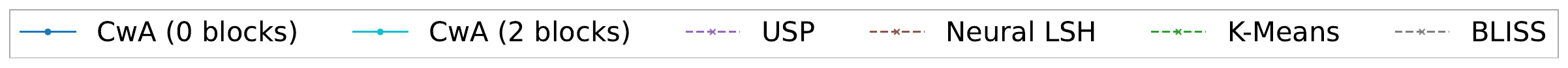}
    \caption{QPS vs.\ recall on 256 clusters ($\nindex=1\,048\,576$):
             Deep, SIFT, Text-to-Image, LAION.}
    \label{fig:qps_256}
\end{figure}

\begin{figure}[H]
    \centering
    \begin{subfigure}[b]{0.24\linewidth}
        \includegraphics[height=3.3cm]{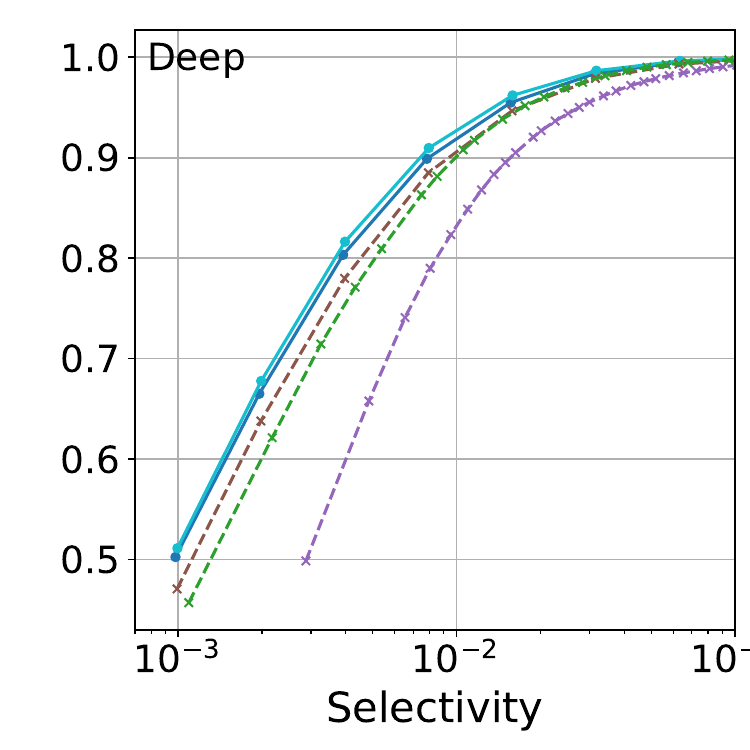}
    \end{subfigure}\hfill
    \begin{subfigure}[b]{0.24\linewidth}
        \includegraphics[height=3.3cm]{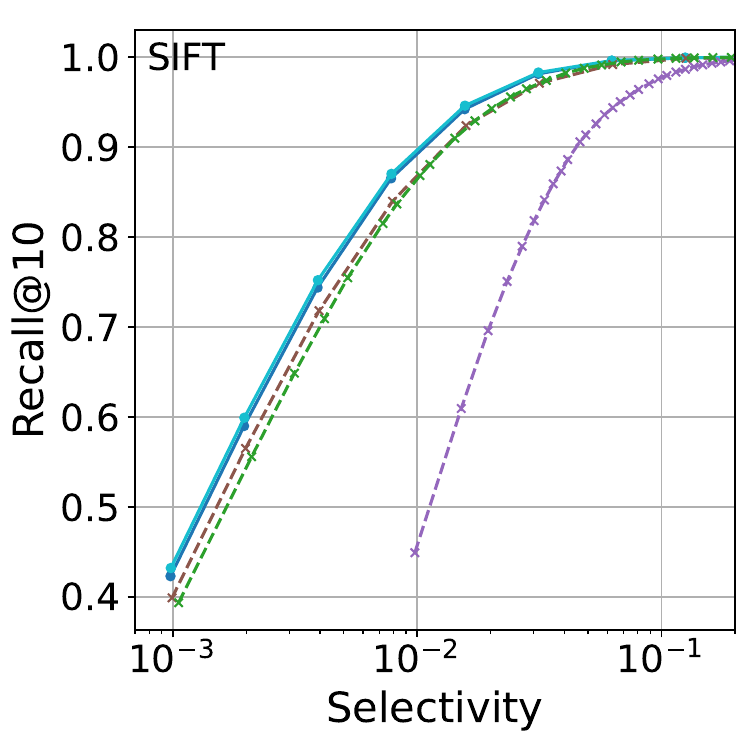}
    \end{subfigure}\hfill
    \begin{subfigure}[b]{0.24\linewidth}
        \includegraphics[height=3.3cm]{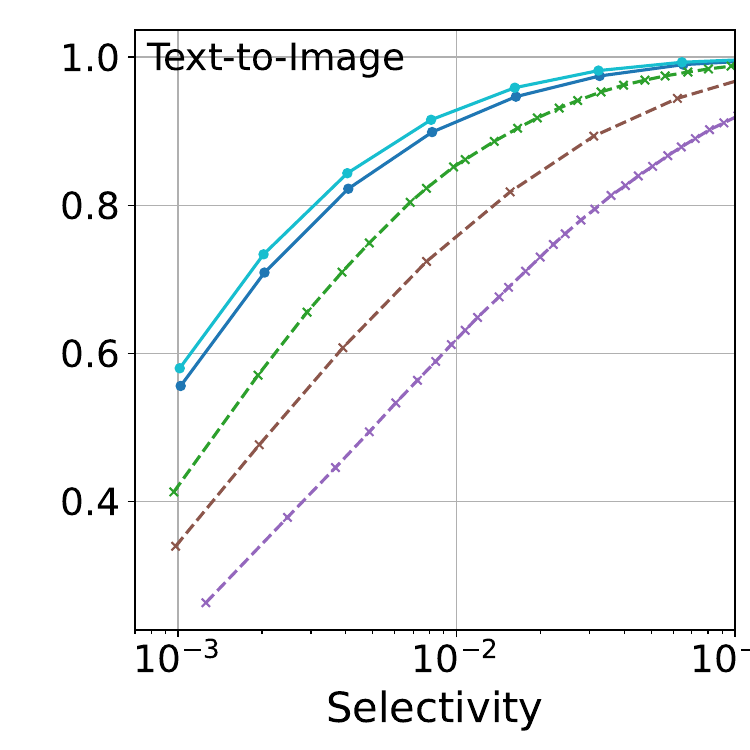}
    \end{subfigure}\hfill
    \begin{subfigure}[b]{0.24\linewidth}
        \includegraphics[height=3.3cm]{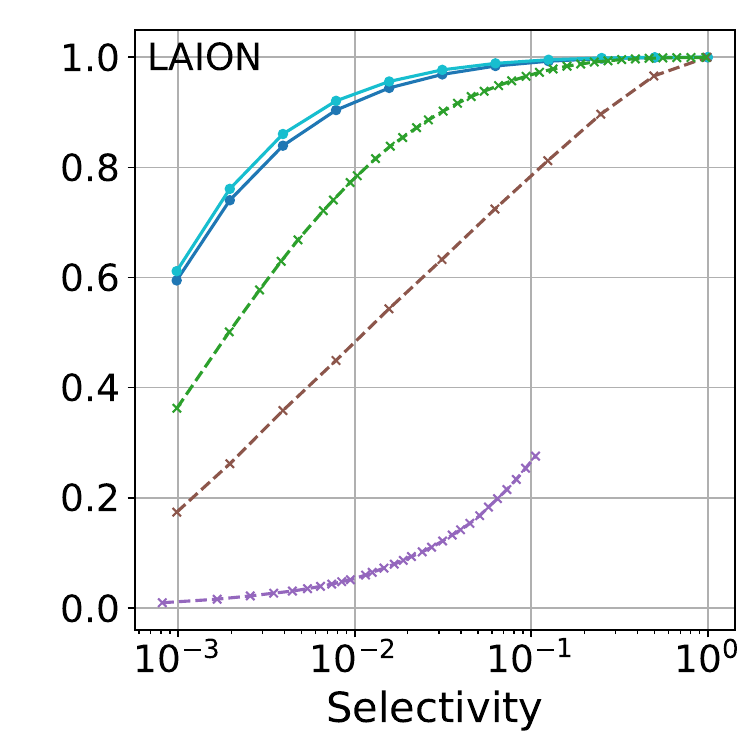}
    \end{subfigure}
    \\[\smallskipamount]
    \includegraphics[height=0.57cm]{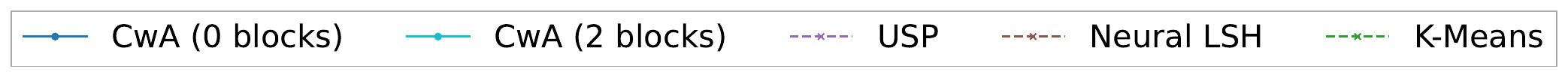}
    \caption{Selectivity vs.\ recall on 1\,024 clusters ($\nindex=1\,048\,576$):
             Deep, SIFT, Text-to-Image, LAION.}
    \label{fig:selectivity_1024}
\end{figure}

\begin{figure}[H]
    \centering
    \begin{subfigure}[b]{0.24\linewidth}
        \includegraphics[height=3.3cm]{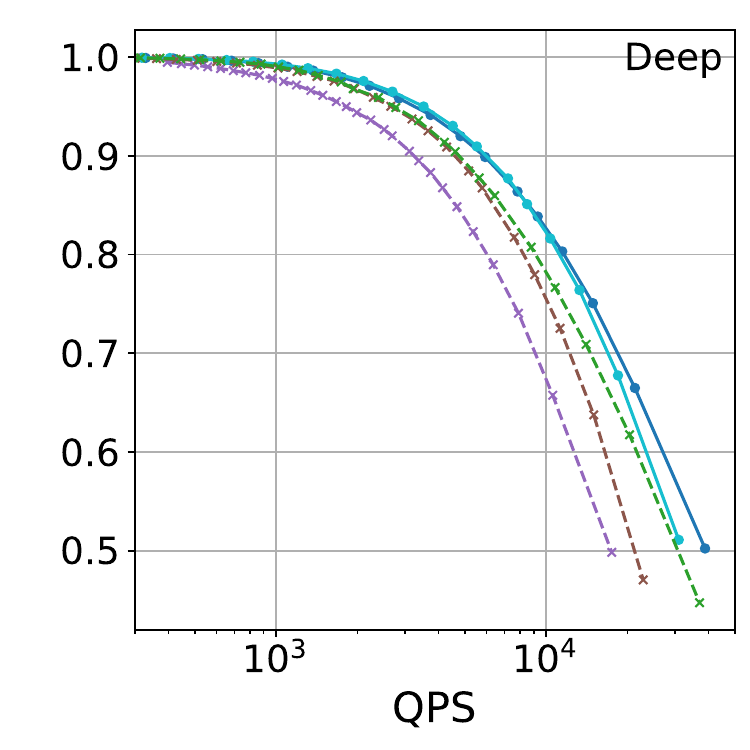}
    \end{subfigure}\hfill
    \begin{subfigure}[b]{0.24\linewidth}
        \includegraphics[height=3.3cm]{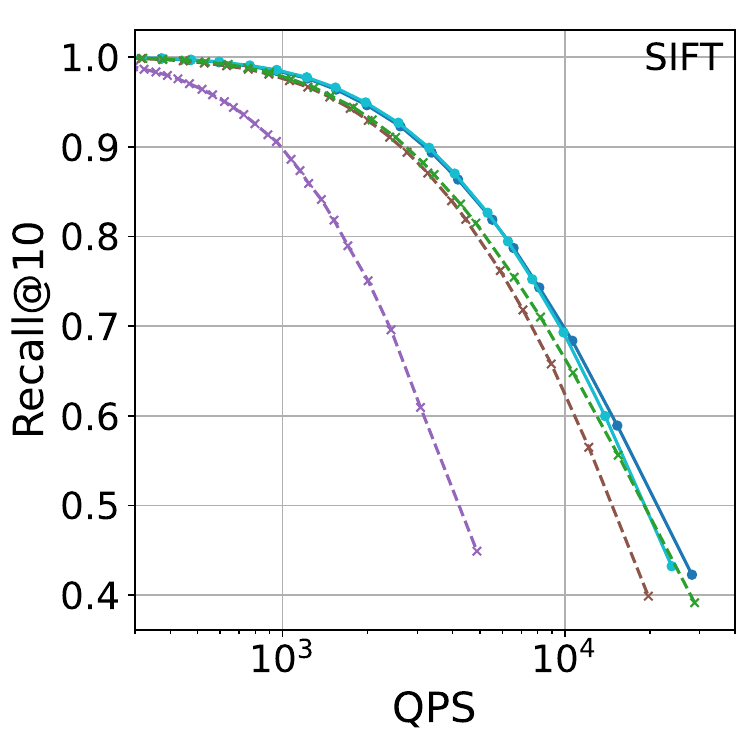}
    \end{subfigure}\hfill
    \begin{subfigure}[b]{0.24\linewidth}
        \includegraphics[height=3.3cm]{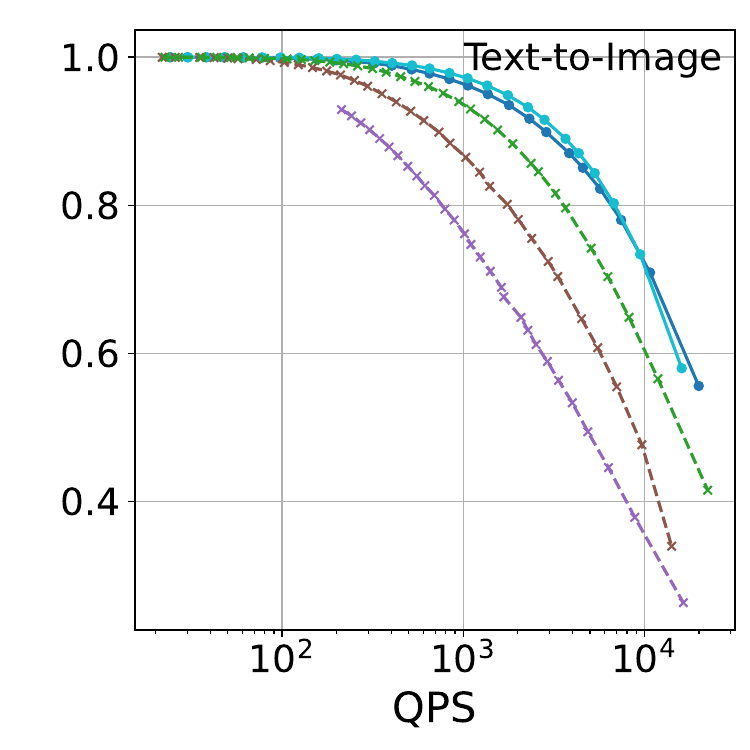}
    \end{subfigure}\hfill
    \begin{subfigure}[b]{0.24\linewidth}
        \includegraphics[height=3.3cm]{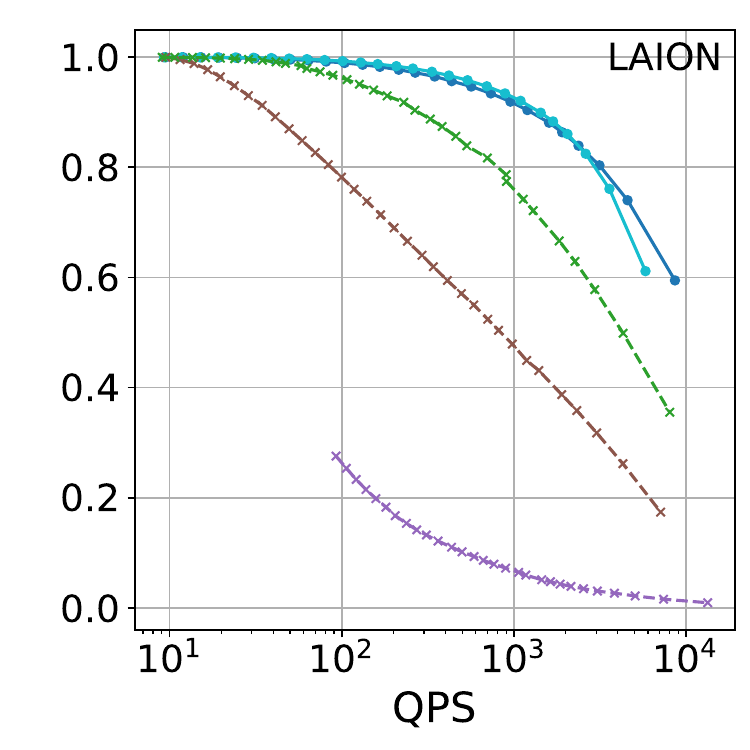}
    \end{subfigure}
    \\[\smallskipamount]
    \includegraphics[height=0.57cm]{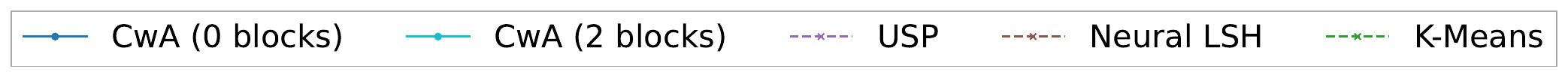}
    \caption{QPS vs.\ recall on 1\,024 clusters ($\nindex=1\,048\,576$):
             Deep, SIFT, Text-to-Image, LAION.}
    \label{fig:qps_1024}
\end{figure}

\begin{figure}[H]
    \centering
    \begin{subfigure}[b]{0.24\linewidth}
        \includegraphics[height=3.3cm]{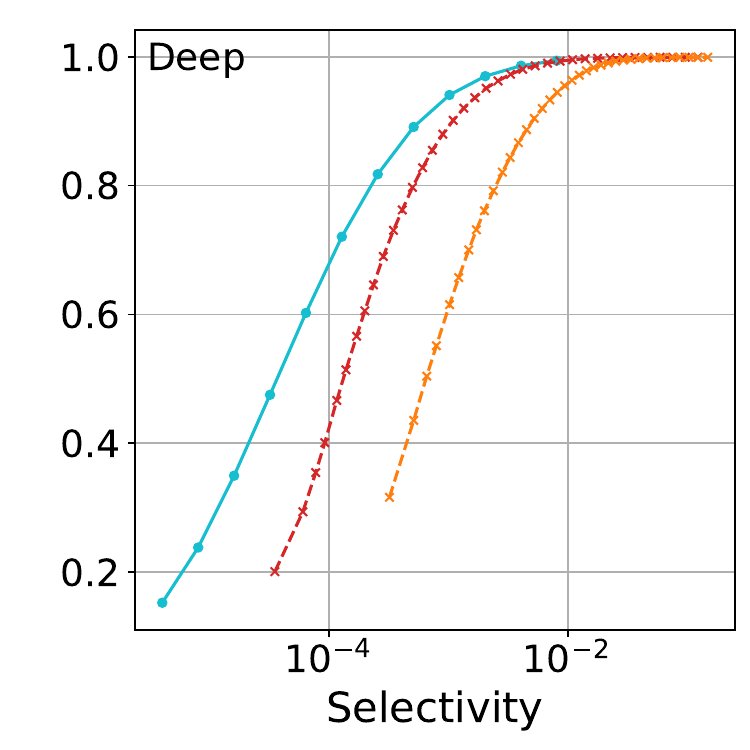}
    \end{subfigure}\hfill
    \begin{subfigure}[b]{0.24\linewidth}
        \includegraphics[height=3.3cm]{figs/plots_qps_recall/plot_selectivity_sift1b_n_index100000000_n_clusters262144.pdf}
    \end{subfigure}\hfill
    \begin{subfigure}[b]{0.24\linewidth}
        \includegraphics[height=3.3cm]{figs/plots_qps_recall/plot_selectivity_text2image10M_n_index100000000_n_clusters262144.pdf}
    \end{subfigure}\hfill
    \begin{subfigure}[b]{0.24\linewidth}
        \includegraphics[height=3.3cm]{figs/plots_qps_recall/plot_selectivity_laion_n_index100000000_n_clusters262144.pdf}
    \end{subfigure}
    \\[\smallskipamount]
    \includegraphics[height=0.57cm]{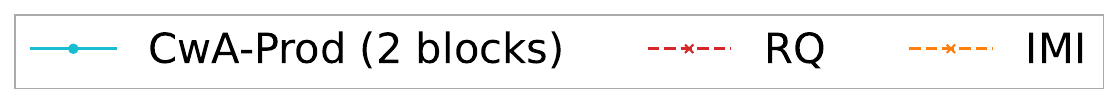}
    \caption{Selectivity vs.\ recall on 262\,144 clusters ($\nindex=100\,000\,000$):
             Deep, SIFT, Text-to-Image, LAION.}
    \label{fig:selectivity_262144_2}
\end{figure}

\begin{figure}[H]
    \centering
    \begin{subfigure}[b]{0.24\linewidth}
        \includegraphics[height=3.3cm]{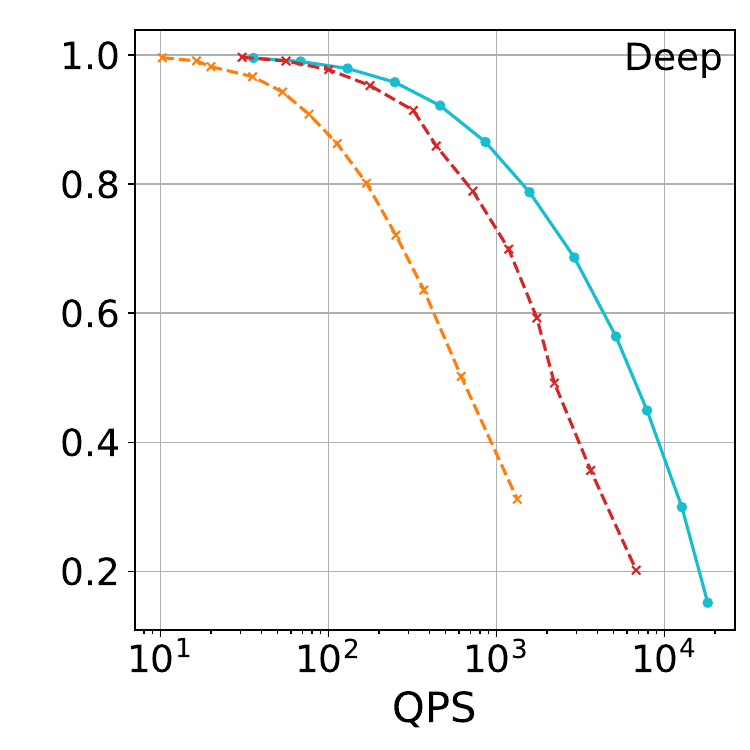}
    \end{subfigure}\hfill
    \begin{subfigure}[b]{0.24\linewidth}
        \includegraphics[height=3.3cm]{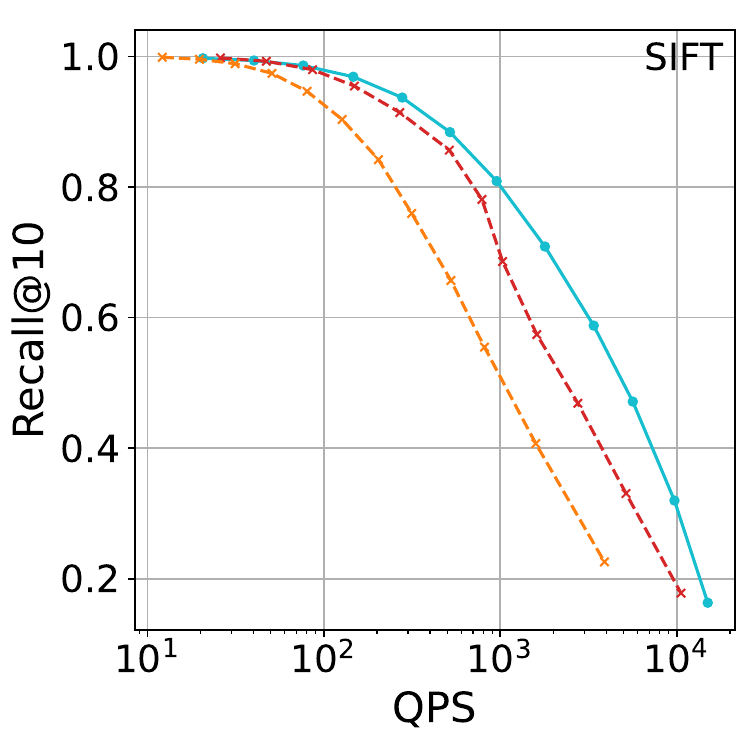}
    \end{subfigure}\hfill
    \begin{subfigure}[b]{0.24\linewidth}
        \includegraphics[height=3.3cm]{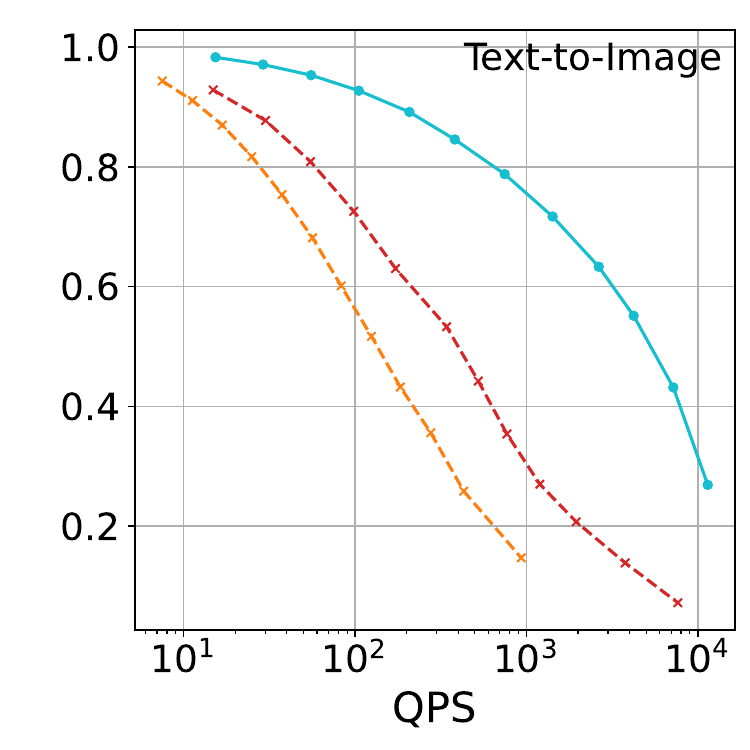}
    \end{subfigure}\hfill
    \begin{subfigure}[b]{0.24\linewidth}
        \includegraphics[height=3.3cm]{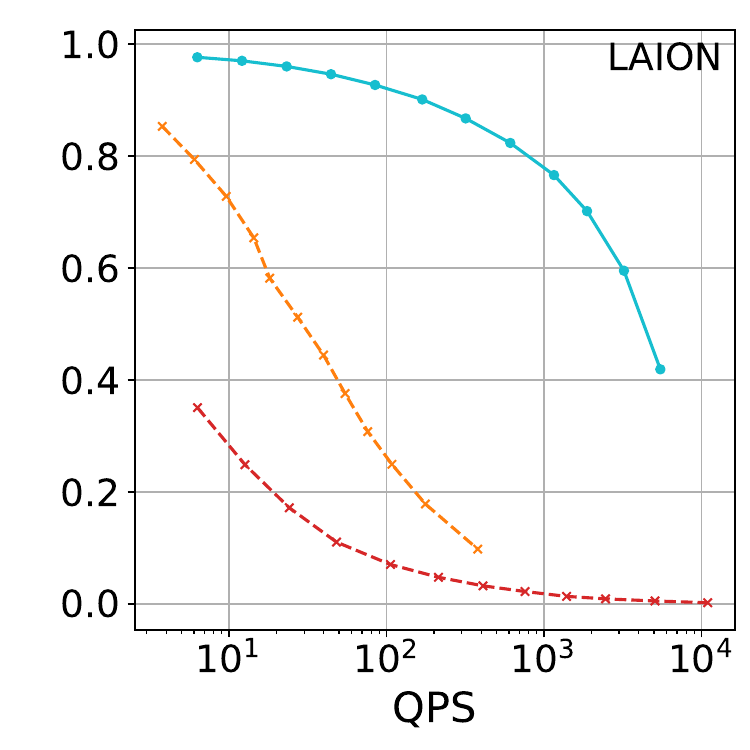}
    \end{subfigure}
    \\[\smallskipamount]
    \includegraphics[height=0.57cm]{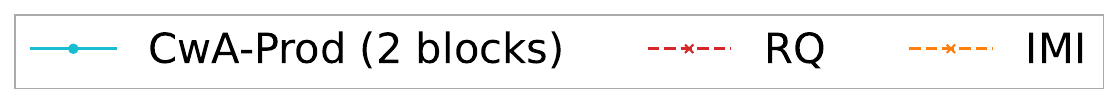}
    \caption{QPS vs.\ recall on 262\,144 clusters ($\nindex=100\,000\,000$):
             Deep, SIFT, Text-to-Image, LAION.}
    \label{fig:qps_262144}
\end{figure}

\begin{figure}[H]
    \centering
    \begin{subfigure}[b]{0.32\linewidth}
        \includegraphics[height=3.3cm]{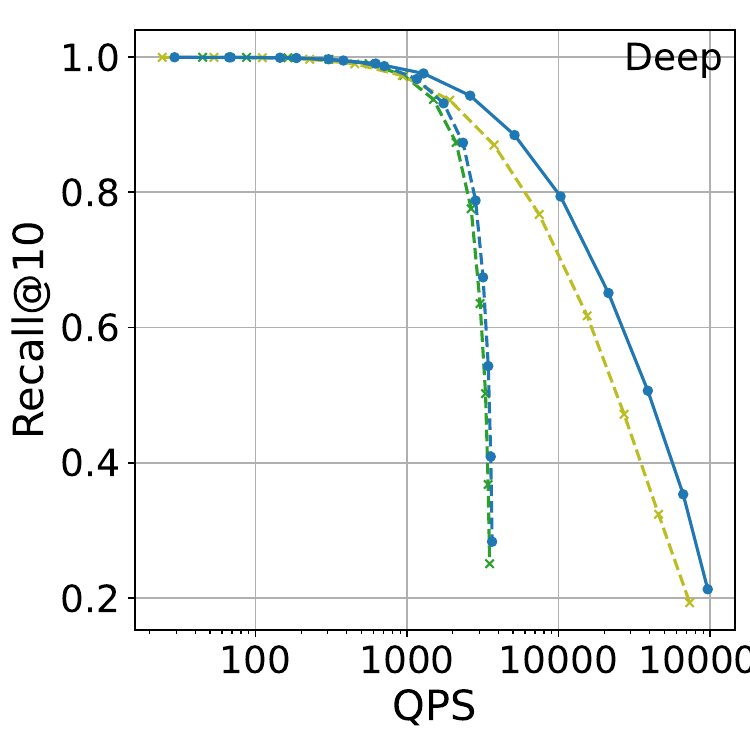}
    \end{subfigure}\hfill
    \begin{subfigure}[b]{0.32\linewidth}
        \includegraphics[height=3.3cm]{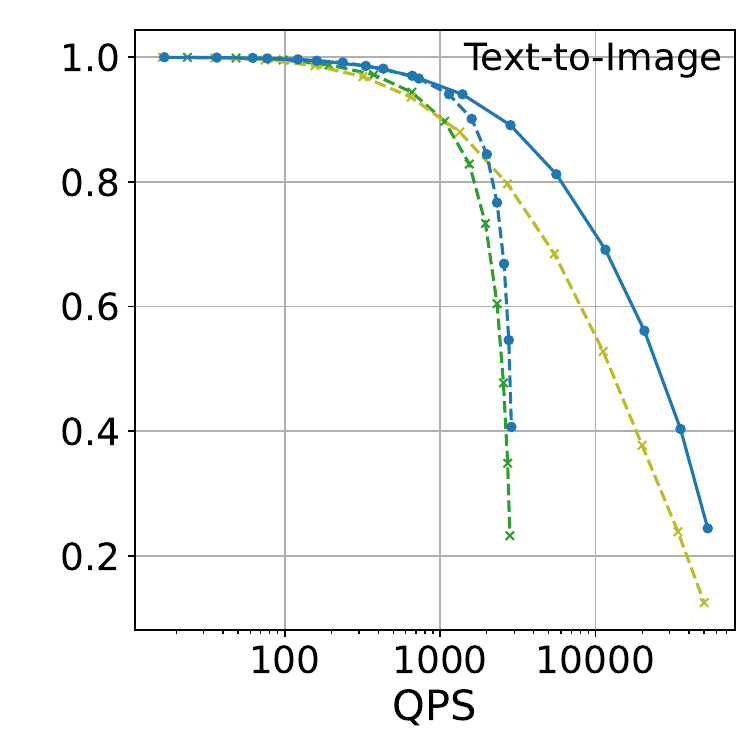}
    \end{subfigure}\hfill
    \begin{subfigure}[b]{0.32\linewidth}
        \includegraphics[height=3.3cm]{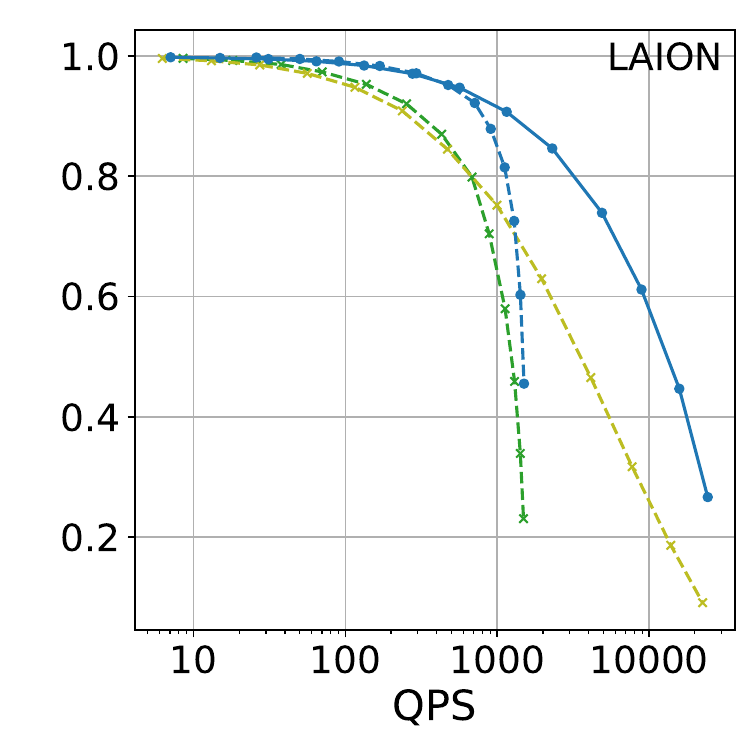}
    \end{subfigure}
    \\[\smallskipamount]
    \includegraphics[height=0.57cm]{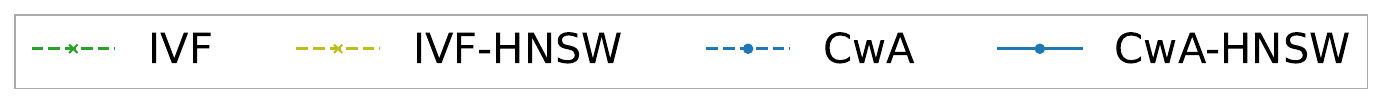}
    \caption{QPS vs.\ recall on 65\,536 clusters ($\nindex=10\,000\,000$):
             Deep, Text-to-Image, LAION}
    \label{fig:qps_65536}
\end{figure}

In this section, we provide the Selectivity (Figures \ref{fig:selectivity_256}, \ref{fig:selectivity_1024} and \ref{fig:selectivity_262144_2}) and QPS (Figures \ref{fig:qps_256}, \ref{fig:qps_1024}, \ref{fig:qps_65536} and \ref{fig:qps_262144}) curves in our experiments. Altogether, they show that \OURS outperforms the baselines in a wide array of settings, setting a new state of the art in space partitioning both for ID and OOD datasets.

Figures \ref{fig:selectivity_256} and \ref{fig:qps_256} show our results for the setting with $\vert C\vert=256$ clusters and $\nindex=1$M database vectors. This setting corresponds to small index scales in Vector Search, but it is the only scale at which all baselines were evaluated in their original publications. The selectivity-recall curves highlight that \OURS consistently achieves better quality partitions on the whole range of recall. We achieve the most impressive improvements on out-of-distribution datasets Text-to-Image and LAION, where neural baselines collapse. At Recall@10$=$0.8, \OURS (2 FFN blocks) is $2.4\times$ faster than K-Means on Text-to-Image and $4.7\times$ faster on LAION. We highlight that \OURS with 0 FFN blocks, a linear model, outperforms baselines using deep neural networks (Neural LSH, USP, BLISS) on all 4 datasets. The gains in selectivity translate to better throughput, as shown in the QPS plots of Figure \ref{fig:qps_256}. In most cases, \OURS with 2 FFN blocks achieves the best QPS at fixed recall. 

Figures \ref{fig:selectivity_1024} and \ref{fig:qps_1024} represent the selectivity and QPS results for the setting with $\vert C\vert=1024$ clusters and $\nindex=1$M database vectors. Again, \OURS consistently outperforms the baselines, most strikingly on out-of-distribution datasets. Interestingly, at high throughput the linear variant (0 FFN blocks) can outperform the deeper one (2 FFN blocks). Since the linear model's forward pass is faster, it pays off when the probing function is the computational bottleneck — i.e.\ at high QPS where few clusters are visited. The effect is clearly visible on Deep: at Recall@10$=$0.6, \OURS (0 blocks) reaches 28\,249 QPS vs.\ 24\,261 for the 2-block variant ($1.16\times$), while at higher recall the deeper model recovers its partition-quality advantage.

Figures \ref{fig:selectivity_262144_2} and \ref{fig:qps_262144} represent the selectivity and QPS results for \PROD with $\vert C\vert=262\,144$ clusters and $\nindex=100$M database vectors. This setting is where we achieve the best relative improvements compared to the baselines. At Recall@10$=$0.8, \PROD achieves $2.1\times$ the throughput of RQ on Deep and $1.4\times$ on SIFT. The advantage grows sharply with distribution shift: on Text-to-Image, \PROD achieves 671 QPS vs.\ 60 for RQ ($11\times$); on LAION, the baselines essentially collapse while \PROD still reaches 834 QPS. 

Figure \ref{fig:qps_65536} shows the QPS results for \HNSW, compared to baselines IVF HNSW and classic IVF (without HNSW on centroids). First, we find that classic IVF is outperformed by \HNSW and IVF HNSW on most recalls, because the HNSW accelerates the centroid search. Second, because it relies on a better quality partition, \HNSW outperforms IVF HNSW in all cases. At Recall@10$=$0.8, \HNSW achieves $1.6\times$ the throughput of IVF-HNSW on Deep (9\,954 vs.\ 6\,271 QPS), $2.3\times$ on Text-to-Image, and $5\times$ on LAION — again with the largest gains on OOD datasets.

\section{Detailed alternating optimization algorithm}
\label{app:altern_opt_detailed}

Algorithm~\ref{alg:altern_opt_detailed} provides the full procedure combining both steps of our alternating optimization procedure.

\begin{algorithm}[H]
\caption{Alternated optimization of $f_{\theta}$ and $h$}
\label{alg:altern_opt_detailed}
\begin{algorithmic}[1]
\Require Training set $\mathcal{D}_{\text{train}} = \{(q_i, \mathcal{N}_{k'}(q_i))\}_{i=1}^{\ntrain}$,
         database vectors $X = (x_{\ell})_{\ell=1}^{\nindex}$,
         number of clusters $|C|$,
         cluster quota $Q$,
         number of alternation steps $A$,
         number of training batches $T_{\text{iter}}$ and $T_{\text{final}}$,
         learning rate $\eta$,
         auction precision $\epsilon > 0$
\State Initialize $f_{\theta}$ randomly
\State Initialize $h$ (with K-Means assignments for \OURS and PQ assignments for \PROD)
\For{$a = 1$ to $A$}
  \State Fix $h$, train $f_{\theta}$ on $\mathcal{D}_{\text{train}}$ for $T_{\text{iter}}$ batches
         by minimizing $\sum_{i=1}^{\ntrain} \mathrm{CE}(p_{h,q_i}, f_\theta(q_i))$
         via back-propagation with learning rate $\eta$
  \State Fix $f_{\theta}$, compute score matrix $S' \in \mathbb{R}^{\nindex \times |C|}$ using Proposition~\ref{prop:fast-score}
  \State Update $h$ by solving $\max_{h} \sum_{\ell=1}^{\nindex} S'_{\ell, h(\ell)}$
         s.t.\ $|\{\ell : h(\ell) = c\}| \leq Q,\ \forall c \in C$,
         via the auction algorithm (Appendix~\ref{app:auction_algo})
\EndFor
\State Fix $h$, train $f_{\theta}$ on $\mathcal{D}_{\text{train}}$ for $T_{\text{final}}$ batches
\State \Return $f_{\theta}, h$
\end{algorithmic}
\end{algorithm}

\section{Additional experimental details}
\label{app:exp_details}
\subsection{Training times}
\label{app:training_times}

We provide here the training times of \OURS across several settings in Table \ref{tab:training_time_cwa_breakdown}. 
At large scale, the neighborhood graph construction is dominant. 

\begin{table}[H]
\centering
\caption{Timing breakdown across configurations on Deep dataset. All timings for index sizes 10M and 100M were obtained on an H100 node, those for 1M were recorded on one GPU.}
\begin{tabular}{l|ccc|c}
\toprule
 & \multicolumn{3}{c|}{\textbf{\FLAT}} & \textbf{\PROD} \\
\midrule
\begin{tabular}[c]{@{}l@{}}\textbf{Database size}\\\textbf{Number of clusters}\end{tabular} & 
\begin{tabular}[c]{@{}c@{}}1M\\256\end{tabular} & 
\begin{tabular}[c]{@{}c@{}}1M\\1024\end{tabular} & 
\begin{tabular}[c]{@{}c@{}}10M\\16k\end{tabular} & 
\begin{tabular}[c]{@{}c@{}}100M\\262k\end{tabular} \\
\midrule
k-NN graph construction            & 2m37s   & 2m37s   & 12m51s  & 7h3m     \\
Backpropagation time               & 21.6s   & 29.9s   & 3m1s    & 12m18s   \\
Auction (score matrix calculation) & 21.9s   & 20.3s   & 4m52s   & 15m16s   \\
Auction (bidding phase)            & 5m11s   & 3m28s   & 17m3s   & 38m45s   \\
\midrule
\textbf{Total}                     & 8m32s   & 6m55s   & 37m46s  & 8h9m     \\
\bottomrule
\end{tabular}
\label{tab:training_time_cwa_breakdown}
\end{table}
We compare \OURS's training times with the baselines in Table \ref{tab:training_time_cwa_compare}.
Our training time is better or similar to methods with competitive accuracy (USP and Neural LSH), but remains clearly above K-Means.
\begin{table}[H]
\centering
\caption{Comparison of training times on GPU: Deep dataset, $\nindex$=1M, 256 clusters }
\begin{tabular}{lccccc}
\toprule
\textbf{Metric} & \textbf{\OURS} & \textbf{Neural LSH} & \textbf{USP} & \textbf{BLISS} & \textbf{K-Means} \\
\midrule
KNN Graph Construction Time & 2m37s & 28.8s & 34.7s & 52.1s & 0 \\
Training Time & 5m54s & 6h50m & 5m12s & 27.5s & 3.2s \\
\bottomrule
\end{tabular}
\label{tab:training_time_cwa_compare}
\end{table}

\subsection{Search time breakdown}
Tables \ref{tab:breakdown_recall2} and \ref{tab:breakdown_recall} break down the search time between cluster selection (Cluster assignment) and actual distance computations with the vectors in selected clusters (Dataset search). Results from Table~\ref{tab:breakdown_recall2} compare \HNSW to its baselines, while Table~\ref{tab:breakdown_recall} compares multi-codebook methods \PROD, IMI and RQ.

\begin{table}[H]
    \centering
    \caption{Breakdown at Recall 0.9, SIFT, $\nindex$=10M, 65536 clusters}
    \begin{tabular}{l|cccc}
        \toprule
        & \textbf{CwA-HNSW} & \textbf{CwA} & \textbf{IVF} & \textbf{IVF-HNSW} \\
        \midrule
        QPS & 2,864 & 1,786 & 1,658 & 2,455 \\
        Cluster search (ms) & 0.09 & 0.25 & 0.26 & 0.09 \\
        Dataset search (ms) & 0.31 & 0.31 & 0.37 & 0.36 \\
        \bottomrule
    \end{tabular}
    \label{tab:breakdown_recall2}
\end{table}

\begin{table}[H]
    \centering
    \caption{Breakdown at Recall 0.6, SIFT, $\nindex$=100M, 262144 clusters}
    \begin{tabular}{l|ccc}
        \toprule
        & \textbf{PROD-CwA} & \textbf{RQ} & \textbf{IMI} \\
        \midrule
        QPS & 3,221 & 1,475 & 685.48 \\
        Cluster search (ms) & 0.06 & 0.04 & 0.01 \\
        Dataset search (ms) & 0.27 & 0.66 & 1.52 \\
        \bottomrule
    \end{tabular}
    \label{tab:breakdown_recall}
\end{table}

\subsection{Parameter counts and forward pass costs}
\label{app:param_count}

Table~\ref{tab:forward_pass} breaks down parameter counts and forward times for \OURS and the baselines. We highlight that \OURS (0 blocks) has the fewest parameters --- as many as K-Means --- but outperforms all baselines in our experiments.

\begin{table}[H]
\centering
\setlength{\tabcolsep}{3.5pt}
\begin{tabular}{lcccccc}
\hline
\textbf{Clusters} & \textbf{Metric} & \textbf{\FLAT} & \textbf{\FLAT} & \textbf{Neural} & \textbf{USP} & \textbf{K-Means} \\
& & \textbf{(0 blocks)} & \textbf{(2 blocks)} & \textbf{LSH} & & \textbf{} \\
\hline
\multirow{2}{*}{\textbf{256}} & Time (µs) & 1.65 & 7.59 & 13.9 & 4.17 & 1.62 \\
& \# param. ($\times 10^3$) & 24.6 & 135.6 & 709.4 & 45.1 & 24.6 \\
\hline
\multirow{2}{*}{\textbf{1024}} & Time (µs) & 2.85 & 8.50 & 18.7 & 7.16 & 2.87 \\
& \# param. ($\times 10^3$) & 98.3 & 209.2 & 1103 & 143.3 & 98.3 \\
\hline
\end{tabular}
\caption{Forward pass time per query (µs) and number of parameters, Deep ($\nindex$=1M), with a batch size of 128. For K-Means, the time corresponds to the distance computation between the query and the centroids.}
\label{tab:forward_pass}
\end{table}

\subsection{Implementation details for baselines}
\label{app:baselines_details}
We provide implementation details for all baselines:
\begin{itemize}
    \item K-Means: We use FAISS's K-Means implementation, applied for 30 iterations. The same algorithm was applied for Residual Quantization~\citep{chen2010approximate} and the Inverted Multi-Index~\citep{imi_babenko}.
    \item Neural LSH: We use the architecture described in Dong et al.~\citep{dong2020learningspacepartitionsnearest}, an MLP with 3 hidden layers of size 512. Training parameters are kept the same as in the original paper. For graph partitioning with KaHiP~\citep{kahip}, we chose the "eco" configuration because the "strong" configuration did not complete after 72 hours on the smallest datasets.
    \item USP: We use the architecture described in Fahim et al.~\citep{fahim2022unsupervisedspacepartitioningnearest}, an MLP with 1 hidden layer of size 200. Training parameters are kept the same as in the original paper.
    \item BLISS: We use the architecture described in Gupta et al.~\citep{Gupta2022BLISSAB}, an MLP with 1 hidden layer of hidden size 512. Training parameters were kept the same as in the original paper. To choose parameter $K$ (see Gupta et al.~\citep{Gupta2022BLISSAB}), we swept on powers of 2 and chose the value with best recall.
\end{itemize}

\section{HNSW plugged on \OURS: CwA-HNSW}
\label{app:hnsw_plugged}
When plugging a graph index on the last linear layer of \OURS, the goal is to retrieve the top-$m$ scoring clusters without computing all scores. Denoting $q'$ the input of the last linear layer and $m_1,\dots,m_{|C|}$ its rows, the scores are dot products $t_i = q' \cdot m_i$, making this a Maximum Inner-Product Search (MIPS) problem. We empirically found that graph-based MIPS (ip-NSW~\citep{NEURIPS2018_229754d7_ipnsw}) was particularly slow in our setting and did not yield the desired speedup.

We therefore replace the dot-product layer with an equivalent formulation whose top-$m$ search reduces to Euclidean nearest neighbor search, for which HNSW is highly efficient. 
The key identity is:
$$\arg\max_i \, \bigl(-\|q' - m_i\|^2\bigr) = \arg\min_i \, \|q' - m_i\|^2,$$
so if scores are negative squared L2 distances, finding the highest scores is exactly an NNS problem on $\{m_i\}$. We therefore reparametrize the scores in the output layer as:
$$t_i = -e^\lambda \cdot \|q' - m_i\|^2,$$
where $\lambda$ is a learnable scalar shared across all clusters, optimized by back-propagation alongside the rows $m_i$. It acts as a global temperature controlling the sharpness of the resulting softmax distribution.

We can thus build an HNSW index on $\{m_i\}_{i=1}^{|C|}$ and query it with $q'$, retrieving the approximate top-$m$ clusters in sublinear time. We found this yielded considerable speedup over the MIPS approach.

\section{Auction algorithm}
\label{app:auction_algo}

This section provides implementation details for the auction algorithm, a cornerstone of \OURS. 

\subsection{Pseudo-code}
\label{app:pseudocode}
We present here the pseudo codes for the auction algorithm. The vanilla auction algorithm as first introduced by Bertsekas~\citep{bertsekas1988auction}, which maps a given number of workers to a higher number of tasks, is described in Algorithm \ref{alg:auction-vanilla}. In our experiments, we use a variant called the auction algorithm for similar objects~\citep{Bertsekas1989AuctionTransportation} described in Algorithm \ref{alg:auction-quota-global-argmax}.
\begin{algorithm}[tbp]
\caption{Vanilla auction algorithm}
\label{alg:auction-vanilla}
\begin{algorithmic}[1]

\Require
$\begin{aligned}[t]
 &X = (x_{\ell})_{\ell=1}^{\nindex},\quad
  S \in \mathbb{R}^{\nindex \times |C|},\quad
  \epsilon > 0 \\
 &p_c = 0 \;\; \forall\, c \in C, n_{\text{iter}} \in \mathbb{N}
\end{aligned}$

\Ensure
$h : \{1,\dots,\nindex\} \to C$

\State Initialize $h(l) \gets -1$ for all $l$ \Comment{dummy value for unassigned items}

\For{iter = 1, 2, ..., $n_{\text{iter}}$}
    \State Initialize bid lists $B_{c} \gets \emptyset$ for all $c \in C$
    \For{each $\ell$ such that $h(\ell) = -1$}
        \For{each $c \in C$} \label{line:vanilla-benefit1}
            \State Compute net benefit: $b_{\ell,c} \gets S_{\ell,c} - p_c$
        \EndFor \label{line:vanilla-benefit2}
        \State Find $c_\ell = \arg\max_{c} b_{\ell,c}$
        \State Let $v_{\ell} \gets b_{\ell,c_\ell}$,\quad $w_\ell \gets \max_{c \neq c_\ell} b_{\ell,c}$
        \State Compute bid increment: $\delta_{\ell} \gets v_{\ell} - w_{\ell} + \epsilon$
        \State Compute bid value: $\hat{b}_{\ell} \gets p_{c_{\ell}} + \delta_{\ell}$
        \State Append bid: $B_{c_{\ell}} \gets B_{c_{\ell}} \cup \{(\ell, \hat{b}_{\ell})\}$
    \EndFor

    \For{$c \in C$}
        \If{$B_{c} \neq \emptyset$}
            \State Choose $({\ell}^{*}, \hat{b}_{\ell^*}) = \arg\max_{(\ell,\hat{b}_{\ell}) \in B_{c}} \hat{b}_{\ell}$
            \Comment{select item with highest bid on $c$}
            \State $j \gets h^{-1}(c)$
            \If{$j \neq \emptyset$}
                \State $h(j) \gets -1$ \Comment{unassign previous item from slot $c$}
            \EndIf
            \State Assign selected item: $h(\ell^{*}) \gets c$
            \State Update price: $p_c \gets \hat{b}_{\ell^{*}}$
        \EndIf
    \EndFor
\EndFor
\State $\mathcal{U} \gets \{\ell : h(\ell) = -1\}$,\quad $\mathcal{F} \gets C \setminus \mathrm{Im}(h)$
\For{each $\ell \in \mathcal{U}$}
    \State $c^* \gets \arg\max_{c \in \mathcal{F}}\, S_{\ell,c}$;\quad $h(\ell) \gets c^*$;\quad $\mathcal{F} \gets \mathcal{F} \setminus \{c^*\}$
    \Comment{assign remaining vectors to best available cluster}
\EndFor
\end{algorithmic}
\end{algorithm}

\begin{algorithm}[tbph]
\caption{Auction algorithm for similar objects with quotas}
\label{alg:auction-quota-global-argmax}
\begin{algorithmic}[1]
\Require
$\begin{aligned}[t]
 &X = (x_{\ell})_{\ell=1}^{\nindex},\quad
  S \in \mathbb{R}^{\nindex \times |C|},\quad
  Q \in \mathbb{N}^*,\quad
  \epsilon > 0 \\
 &p_{c,s} = 0 \;\; \forall\, c \in C,\, s \in \{1,\dots,Q\}, n_{\text{iter}} \in \mathbb{N} \\
\end{aligned}$
\Ensure
$\begin{aligned}[t]
 &h : \{1,\dots,\nindex\} \to C,\\
 &g : \{1,\dots,\nindex\} \to \{1,\dots,Q\}
\end{aligned}$
\State Initialize $h(\ell) \gets -1$, $g(\ell) \gets -1$ for all $\ell$
\Comment{put dummy values for unassigned indices}
\For{iter = 1, 2, ..., $n_{\text{iter}}$}

    \State Initialize $s^{*}_c \gets \arg\min_{s} p_{c,s}$ for all $c$
    \State Initialize bid lists $B_{c} \gets \emptyset$ for all $c$
    \Comment{store all bids for each cluster}
    \For{each $\ell$ s.t. $h(\ell) = -1$}
        \For{each cluster $c$} \label{line:benefit_computation}
            \State Compute net benefit: $b_{\ell,c} \gets S_{\ell,c} - p_{c,s^{*}_c}$
        \EndFor \label{line:benefit_computation2}
        \State Find $c_{\ell} = \arg\max_{c} b_{\ell,c}$
        \State Let $v_\ell \gets b_{\ell,c_{\ell}}$, \; $w_\ell \gets \max_{c \neq c_{\ell}} b_{\ell,c}$
        \State Compute bid increment: $\delta_{\ell} \gets v_{\ell} - w_{\ell} + \epsilon$
        \State Compute bid value: $\hat{b}_{\ell} \gets p_{c_{\ell},s_{c_{\ell}}^{*}} + \delta_{\ell}$
        \State Append bid: $B_{c_{\ell}} \gets B_{c_{\ell}} \cup \{(\ell, \hat{b}_{\ell})\}$
    \EndFor
    \For{$c \in C$}
        \If{$B_{c} \neq \emptyset$}
            \State Choose $(\ell^{*}, \hat{b}_{\ell^{*}}) = \arg\max_{(\ell,\hat{b}_{\ell}) \in B_{c}} \hat{b}_{\ell}$
            \Comment select item with highest bid
            \State $j \gets (h,g)^{-1}(c,s^{*}_{c})$
            \If{$j \neq \emptyset$}
                \State $h(j) \gets -1$;\quad $g(j) \gets -1$
                \Comment unassign previous item on selected slot
            \EndIf
            \State Assign selected item : $h(\ell^{*}) \gets c$, $g(\ell^{*}) \gets s_{c}^{*}$
            \State Update slot price: $p_{c,s^{*}_c} \gets \hat{b}_{\ell^{*}}$
        \EndIf
    \EndFor
\EndFor
\State $\mathcal{U} \gets \{\ell : h(\ell) = -1\}$,\quad $\mathcal{F} \gets \{(c,s) : (c,s) \notin \mathrm{Im}(h,g)\}$
\For{each $\ell \in \mathcal{U}$}
    \State $(c^*, s^*) \gets \arg\max_{(c,s) \in \mathcal{F}}\, S_{\ell,c}$;\quad $h(\ell) \gets c^*$;\quad $g(\ell) \gets s^*$;\quad $\mathcal{F} \gets \mathcal{F} \setminus \{(c^*, s^*)\}$
    \Comment{assign remaining vectors to best available slot}
\EndFor
\end{algorithmic}
\end{algorithm}

\paragraph{Additional details}
The auction algorithm is executed with $n_{\text{iter}}=10,000$ bidding steps. The coefficient $\epsilon$, which represents the level of approximation in the auction (see Bertsekas~\citep{bertsekas1988auction}), is set to $1 \times 10^{-2}$. If some items remain unassigned at the end of the auction algorithm, we choose to assign them to the most appropriate (highest entry in the score matrix $S$) cluster that's not full (see Algorithms~\ref{alg:auction-vanilla} and \ref{alg:auction-quota-global-argmax}). In practice, they represent a tiny fraction ($<0.1\%$) of the database and do not affect performance.
\subsection{Fast score matrix computation}
\label{app:trick_s}

Computing the score matrix $S$ from Equation~\ref{eq:score_matrix} naively requires evaluating $\log(\hat{p}_{\theta,q_i}(c))$ for every query-cluster pair, which involves a softmax and a $\log$.
We show that the softmax and the $\log$ cancel each other, yielding an equivalent score matrix computed from raw logits only.

\label{prop:fast-score}
Let $z_i$ %
denote the logits before softmax, so that $\hat{p}_{\theta,q_i}(c) = \exp(z_{i,c}) / \sum_{c'}\exp(z_{i,c'})$.
Then for any fixed $\ell$,
\begin{align}
S_{\ell,c}
&= -\frac{1}{k'}\sum_{i=1}^{\ntrain} \log\bigl(\hat{p}_{\theta,q_i}(c)\bigr)\,\mathbf{1}[\ell \in \mathcal{N}_{k'}(q_i)] \nonumber \\
&= \underbrace{-\frac{1}{k'}\sum_{i=1}^{\ntrain} z_{i,c}\,\mathbf{1}[\ell \in \mathcal{N}_{k'}(q_i)]}_{S'_{\ell,c}}
\;+\; \underbrace{\frac{1}{k'}\sum_{i=1}^{\ntrain} \log\!\Bigl(\textstyle\sum_{c'}\exp(z_{i,c'})\Bigr)\,\mathbf{1}[\ell \in \mathcal{N}_{k'}(q_i)]}_{\textrm{independent of } c}. 
\label{eq:decomposed_score_matrix}
\end{align}

The second term does not depend on $c$ and therefore does not affect the assignment $h$.
The modified score $S'_{\ell,c}$ only involves the raw logits $z_{i,c}$, avoiding all log and softmax operations and substantially speeding up the score matrix computation.

\subsection{Sparsification of the score matrix for the auction algorithm}
\label{subsec:sparsification}

The score matrix $S \in \mathbb{R}^{\nindex \times |C|}$ is the main memory bottleneck of the auction algorithm, especially since it must fit in GPU RAM.
We aim at sparsifying $S$ by allowing only a small number $\kappa \ll |C|$ of possible clusters per database vector.

For this, we run a preliminary K-Means clustering on the database vectors $X$, producing $|C|$ centroids.
For each database vector $x_\ell$, we allow only assigning to the $\kappa$ closest centroids.
The score matrix entries $S_{\ell, c}$ are computed and stored only for these $\kappa$ candidate clusters; all other assignments are \emph{forbidden}, \ie conceptually set to $S_{\ell,c} = -\infty$.
This reduces the memory footprint from $\mathcal{O}(\nindex \times |C|)$ to $\mathcal{O}(\nindex \times \kappa)$.
We typically set $\kappa$ at a value between 5 and 10\% of the total number of clusters. Note that this method doesn't apply to \PROD, where we use a different trick described in Appendix \ref{app:tricks_prod}.

\subsection{Memory trick for \PROD}
\label{app:tricks_prod}
In order to scale to a higher number of clusters, we used tricks in our implementation of the auction in \PROD. In this subsection we show how we reduce the memory requirement by factorizing the score matrix in two much smaller ones.

As highlighted in \ref{subsec:architectures}, the scores returned by the model are expressed as: $g_\theta(r, s)_{i, j} = \gamma_{i,j} \times (r_i + s_j)$, where $r, s \in \mathbb{R}^{\sqrt{|C|}}$ are the outputs of the two linear layers for a given query.
Here $r_{p,i}$ and $s_{p,j}$ denote the $i$-th and $j$-th components of the two projection outputs for training query $q_p$, and $z_{p,c}$ denotes the pre-softmax logit for cluster $c$ (see Appendix~\ref{app:trick_s}).
Following Appendix~\ref{app:trick_s}, for cluster $(i,j)$, indexed as $i\times\sqrt{|C|}+j$ with $i,j\in\{1,\ldots,\sqrt{|C|}\}$, the score matrix term relevant to the optimization objective can be expressed as:
\begin{align*}
\forall \ell \in \{1,\dots,\nindex\}, S'_{\ell,\,i\times\sqrt{|C|}+j} &= -\frac{1}{k'}\sum_{p=1}^{\ntrain} z_{p,\,i\times\sqrt{|C|}+j}\,\mathbf{1}[\ell \in \mathcal{N}_{k}(q_p)]
\; \\
&= -\frac{1}{k'}\sum_{p=1}^{\ntrain} \gamma_{i,j}(r_{p,i} + s_{p,j})\,\mathbf{1}[\ell \in \mathcal{N}_{k'}(q_p)]
\; \\
&= \gamma_{i,j}(\underbrace{- \frac{1}{k'} \sum_{p=1}^{\ntrain} r_{p,i}\,\mathbf{1}[\ell \in \mathcal{N}_{k'}(q_p)]}_{S^{1}_{\ell,i}}) + \gamma_{i,j}(\underbrace{-\frac{1}{k'} \sum_{p=1}^{\ntrain} s_{p,j}\mathbf{1}[\ell \in \mathcal{N}_{k'}(q_p)]}_{S^{2}_{\ell,j}}) \\
&= \gamma_{i,j} (S^{1}_{\ell,i} + S^{2}_{\ell,j}).
\; \\
\end{align*}

Instead of keeping the full $S'$ matrix of size $(\nindex,|C|)$ in memory, we only keep the matrices $S^1$ and $S^2$ of size $(\nindex,\sqrt{|C|})$ along with the $\gamma$ matrix of size $(\sqrt{|C|},\sqrt{|C|})$. Rows of $S'$ are then computed on-the-fly at the bidding phase from those of $S^1$ and $S^2$, reducing the memory footprint of the algorithm by a factor of $\sqrt{|C|}/2$. 

\subsection{Efficient top-K computation for \PROD}
\label{app:avxoptim}

There is an efficient algorithm used for the Inverted Multi-Index~\citep{imi_babenko} to compute the top-$k$ values of Equation~\ref{eq:prod} when $\gamma_{i,j}$ is constant.
It is based on a priority queue that is updated by picking from the sorted arrays $r_i$ and $t_j$.
However, this does not apply in our case. The method we used, which proved the most efficient among those experimented, is based on vectorized computing of 16 queries at a time.
The top-$k$ results are then tracked with a reservoir bucket that is also vectorized.

\section{Scaling law}
\label{subsubsec:exp-scaling}

In ANNS applications, the number of available training queries is often much smaller than the index size, making data efficiency a genuine concern.
As explained in Subsection~\ref{subsec:supervision}, we denote by $k'$ the number of nearest neighbors used to construct supervision labels, to distinguish it from the retrieval target $k = 10$.
When training data is scarce, setting $k' > k$ smooths the supervision targets by spreading them across more clusters per query, acting as a regularizer --- despite deviating from the exact Recall@$k$ objective.
This section derives two scaling laws that make these trade-offs precise: one to set $k'$ optimally given $\nindex$ and $\ntrain$, and one to determine how much training data is needed as a function of index size.

We swept over pairs $(\nindex, \ntrain)$ and identified the optimal $k'$ for each via grid search on evaluation recall. The resulting log-linear fit, derived on the Deep dataset for vanilla \OURS (0 FFN blocks), is:
\begin{equation}
\label{eq:k_law}
k' = 0.78 \cdot \nindex^{0.535} \cdot \ntrain^{-0.220}.
\end{equation}

\begin{figure}[H]
    \centering
    \includegraphics[height=5.0cm]{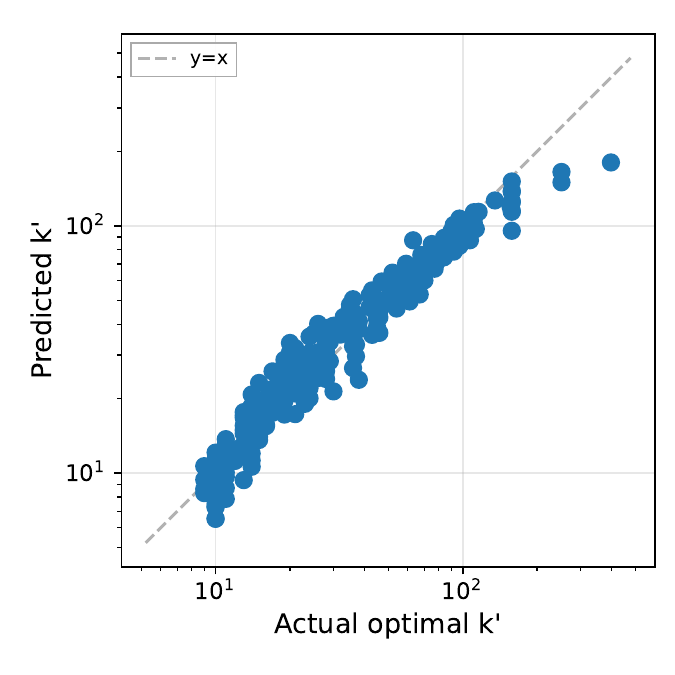}
    \quad
    \includegraphics[height=5.0cm]{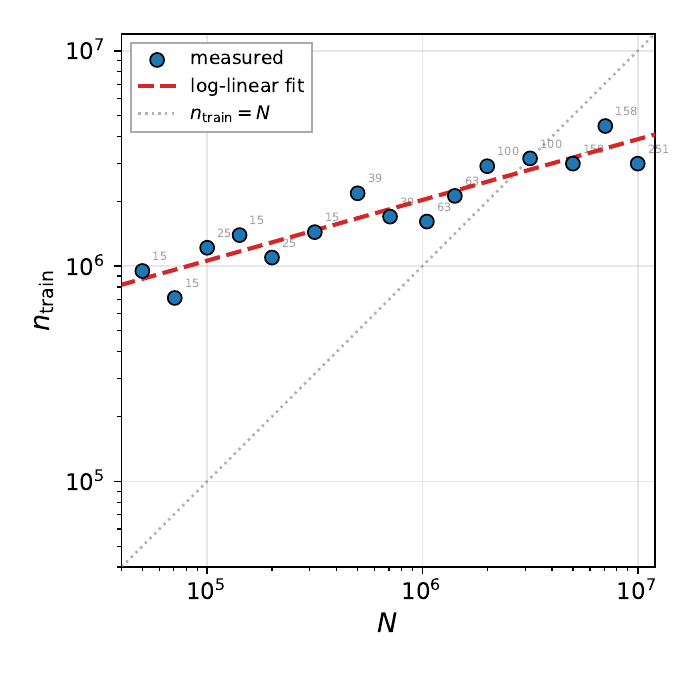}
    \caption{
        \textbf{Left:} Scaling law for the number of nearest neighbors in supervision ($k'$): Predicted vs Actual ($R^2 = 0.938$).
        \textbf{Right:} Scaling law for training data requirements. $\ntrain$ is the minimum training set size to reach 99\% of optimal recall when 1\% of the database is scanned. Grey annotations show the optimal $k'$ at each operating point, determined by grid search.
    }
    \label{fig:scaling_laws}
\end{figure}

The fit achieves $R^2 = 0.938$ (Figure~\ref{fig:scaling_laws}, left). The exponents confirm that $k'$ increases with index size (positive exponent on $\nindex$) and decreases with more training data (negative exponent on $\ntrain$): the more data-constrained the setting, the wider the supervision neighborhood should be for a fixed recall target.

A more practically useful result is the scaling law for training data requirements. For a range of index sizes, we computed the minimum $\ntrain$ to reach 99\% of optimal Recall@10 at 1\% scan\footnote{Recall is measured for distance computations on 1\% of the whole dataset}, with $k'$ set according to Equation~\ref{eq:k_law}. The result is strikingly sub-linear (Figure~\ref{fig:scaling_laws}, right):

\begin{equation}
\ntrain = 41331 \cdot \nindex^{0.282} \quad (R^2 = 0.871)
\end{equation}

An exponent of $0.282$ — far below the linear baseline of $1$ — implies that scaling the index by $100\times$ only requires $\approx 4\times$ more training queries, and a $1000\times$ larger index requires only $\approx 7\times$ more. Equivalently, the ratio $\ntrain / \nindex$ shrinks as the index grows: large-scale deployments are the setting where \OURS is \emph{easiest} to train relative to database size. To our knowledge, this is the first such empirical law for training data requirements in learned vector search. Whether this sub-linear scaling behavior generalizes to other learning-based methods for vector search remains open, but we find it an encouraging sign for the data efficiency of learned indexing at scale.

\end{document}